\documentclass[fleqn,usenatbib]{mnras}

\usepackage{newtxtext,newtxmath}

\usepackage[T1]{fontenc}

\DeclareRobustCommand{\VAN}[3]{#2}
\let\VANthebibliography\thebibliography
\def\thebibliography{\DeclareRobustCommand{\VAN}[3]{##3}\VANthebibliography}

\usepackage{graphicx}	
\usepackage{amsmath}	

\newcommand{\ms}{\ifmmode{\,M_{\odot}}\else{\,$M_{\odot}$}\fi}
\newcommand{\kms}{\ifmmode{\mathrm{\,km\,s^{-1}}}\else{\,$\mathrm{km\,s^{-1}}$}\fi}
\newcommand{\teff}{\ifmmode{T_\mathrm{eff}}\else{$T_\mathrm{eff}$}\fi}

\title[A GTC spectroscopic study of three spiders]{A GTC spectroscopic study of three spider pulsar companions: line-based temperatures, a new face-on redback, and improved mass constraints}

\author[J. Simpson et al.]
{Jordan A. Simpson,$^{1}$\thanks{E-mail: jordan.simpson-parry@ntnu.no}
Manuel Linares,$^{1,2}$\thanks{E-mail: manuel.linares@ntnu.no}
Jorge Casares,$^{3,4}$
Tariq Shahbaz,$^{3,4}$
Bidisha Sen,$^{1}$
\newauthor
and Fernando Camilo$^{5}$
\\
$^{1}$Department of Physics, Norwegian University of Science and Technology, H\o gskoleringen 5, NO-7491 Trondheim, Norway\\
$^{2}$Departament de F\'isica, EEBE, Universitat Polit\`ecnica de Catalunya, Av. Eduard Maristany 16, E-08019 Barcelona, Spain\\
$^{3}$Instituto de Astrof\'isica de Canarias, E-38205 La Laguna, Tenerife, Spain\\
$^{4}$Departamento de Astrof\'isica, Universidad de La Laguna, E-38206 La Laguna, Tenerife, Spain\\
$^{5}$South African Radio Astronomy Observatory, ZA-7705 Cape Town, South Africa
}
\date{Accepted 2024 December 6. Received 2024 December 6; in original form 2024 August 20}

\pubyear{2024}

\begin{document}
\label{firstpage}
\pagerange{\pageref{firstpage}--\pageref{lastpage}}
\maketitle

\begin{abstract}
We present GTC-OSIRIS phase-resolved optical spectroscopy of three compact binary MSPs, or `spiders': PSR J1048+2339, PSR J1810+1744, and (for the first time) PSR J1908+2105. For the companion in each system, the temperature is traced throughout its orbit, and radial velocities are measured. The radial velocities are found to vary with the absorption features used when measuring them, resulting in different radial velocity curve semi-amplitudes: for J1048 ($K_\mathrm{metals, red} = 344 \pm 4\kms{}$, $K_\mathrm{metals, blue} = 372 \pm 3\kms{}$) and, tentatively, for J1810 ($K_\mathrm{Balmer} = 448 \pm 19\kms{}$,  $K_\mathrm{metals} = 491 \pm 32\kms{} $). 
With existing inclination constraints, this gives the neutron star (NS) and companion masses $M_\mathrm{NS} = 1.50$--$2.04\ms{}$ and $M_2 = 0.32$--$0.40\ms{}$ for J1048, and $M_\mathrm{NS} > 1.7\ms{}$ and $M_2 = 0.05$--$0.08\ms{}$ for J1810. 
For J1908, we find an upper limit of $K_2 < 32\kms{}$, which constrains its mass ratio $q = M_2 / M_\mathrm{NS} > 0.55$ and inclination $i < 6.0^\circ$, revealing the previously misunderstood system to be the highest mass ratio, lowest inclination redback yet. This raises questions for the origins of its substantial radio eclipses.
Additionally, we find evidence of asymmetric heating in J1048 and J1810, and signs of metal enrichment in J1908. 
We also explore the impact of inclination on spectroscopic temperatures, and demonstrate that the temperature measured at quadrature ($\phi = 0.25, 0.75$) is essentially independent of inclination, and thus can provide additional constraints on photometric modelling.

\end{abstract}



\section{Introduction}
\label{sec:intro}

Millisecond pulsars (MSPs) are extremely rapidly-rotating ($P_\mathrm{s}<$ 30 ms) neutron stars, spun-up via accretion from a companion star. Thanks to the \textit{Fermi} Large Area Telescope (\textit{Fermi}-LAT), along with continually improving detection techniques \citep{fermi23}, their population continues to grow at a rapid pace -- many of these systems emit as GeV-bright sources to which \textit{Fermi} is particularly sensitive. Targeted follow-up of unassociated \textit{Fermi}-LAT sources with radio and optical telescopes has been especially fruitful in discovering new MSPs.

Black widows and redbacks, with binary periods $\lesssim 1$ day, collectively referred to as `spiders', form a crucial subset of these systems and are known to contain some of the most massive neutron stars \citep{strader19, linares20, romani22}. 
Many of these compact binary MSPs are located relatively nearby (within a few kpc) and out of the galactic plane, making them well suited to spectroscopic campaigns from 8--10 m ground-based optical telescopes. Such studies are not only capable of determining precise neutron star masses, but can also measure and constrain the effects of binary interactions present in these systems, such as strong irradiation by the  pulsar wind, intrabinary shocks, asymmetric heating, and supernova enrichment \citep{voisin20, shahbaz22}.

To determine precise neutron star masses, the orbital inclination, $i$, must be well-constrained. This is typically done by modelling optical light curves. Unfortunately, this often has issues with degeneracies between model parameters, such as companion temperature, extinction, and distance. Spectroscopic temperature measurements can help lift these degeneracies, resulting in more robust orbital inclination measurements and, by extension, neutron star masses.

In addition, the radial velocity semi-amplitude of the companion ($K_2$) must be precisely measured. The effects of irradiation have to be accounted for, as this can shift the centre of light of a given absorption feature significantly from the centre of mass of the companion. Applying a `$K$-correction' can account for this effect \citep{wade88}, but introduces significant uncertainties -- especially on the neutron star mass as $M_\mathrm{NS} \propto (K_2)^3$. 
Instead, by carefully measuring the absorption features associated with the hot and cold sides of the companion, we can correct for the effects of irradiation and constrain the `true' $K_2$. Thus we obtain an `empirical $K$-correction', as introduced by \citet{linares18}.
A similar approach is applied here.

Here we report the results of the analysis of optical spectroscopy of the following three spider systems:

\textbf{PSR J1048+2339} (hereafter J1048) is a redback system, with an $M_\mathrm{2, min} = 0.30\ms{}$ companion in a 6.0-h orbit with a $P_\mathrm{s} = 4.66$ ms radio pulsar (where $M_\mathrm{2, min}$ is the minimum companion mass, calculated by assuming a canonical neutron star mass of $1.4\ms{}$ and $i=90^\circ$) . The pulsar was initially discovered by Arecibo follow-up of a \textit{Fermi} source by \citet{cromartie16}, where it was classified as a redback due to its high minimum companion mass and substantial radio eclipses, and was determined to be relatively nearby, with a YMW16 \citep{yao17} dispersion measure (DM) distance of 1.7 kpc.
Subsequent multi-wavelength observations by \citet{deneva16} confirmed the redback nature of the system and found optical modulations characteristic of pulsar-wind irradiation, with potential evidence of asymmetric heating. Optical photometry by \citet{yap19} later revealed the ``face changing'' nature of the companion, which presented a rapid transition between a strongly irradiated state, with a single-peaked light curve, and a weakly irradiated state, with a double-peaked light curve dominated by ellipsoidal modulation (i.e. due to the tidal distortion of the companion). \citet{zanon21} found bright and highly variable H$\alpha$ emission in the system, interpreted as an effect of the intrabinary shock, and constrained $K_2$ to be between 291 and 348\kms{}.

To supplement the observations presented here, simultaneous photometry of J1048 was obtained \citep{tidemann23} which shows the system to be in its irradiated state, and confirms the asymmetry found by \citet{deneva16}. More recently, \citet{clark23} used J1048's gamma-ray eclipses to constrain the orbital inclination to remarkably edge-on, at $i \gtrsim 80^\circ$ -- with the exact limit depending on the Roche-lobe filling factor of the companion.

\textbf{PSR J1810+1744} (hereafter J1810), discovered in a Green Bank Telescope survey of unidentified Fermi sources \citep{hessels11}, is a black widow spider with a degenerate companion of $M_\mathrm{2, min} = 0.045\ms{}$. This heavily-irradiated companion is in a compact, 3.6-h orbit with a rapidly-spinning 1.66-ms pulsar. J1810's faint and highly variable optical counterpart was identified in \textit{Gemini North} photometry by \citet{breton13}, which presented extreme fluctuations of $\Delta g > 3$\,mag due to its tight orbit and highly-energetic pulsar. Further photometric observations at a higher time resolution from \citet{schroeder14} and \citet{romani21} confirm this extreme variability, and give tantalising evidence for asymmetry around maximum light. The latter considers a variety of heating models that could reproduce this effect, deriving a neutron star mass of $2.11\pm0.04\ms{}$ \citep[updated in][]{kandel23} with an inclination of $66.3\pm0.5^\circ$ -- although the exact value varies depending on the model selected. 

\citet{romani21} also obtained spectroscopic observations of J1810 from which radial velocities have been measured. These were marginalised over as part of the photometric fitting process to give a model $K_2$ of $462.9 \pm 2.2\kms{}$ \citep[updated result from][]{kandel23}.

While no gamma-ray eclipses are observed in J1810, the lack thereof allows a maximum inclination of 84.7$^\circ$ to be derived, raising the minimum neutron star mass to 1.59\ms{} \citep{clark23}.

\textbf{PSR J1908+2105} (hereafter J1908) is another spider pulsar discovered in Arecibo searches by \citet{cromartie16}, with a short, 3.5-h orbital period. While its minimum companion mass of $M_\mathrm{2, min} = 0.055\ms{}$ is reminiscent of a black widow, J1908 is classified as a redback due to its substantial radio eclipses covering approximately 40 per cent of its orbit. Preliminary optical results from \citet{beronya23} appear to support this, with a low amplitude ($\sim$0.2\,mag) light curve and an approximate spectral classification of K to early M. Alternatively, \citet{deneva21} suggest J1908 could be representative of an intermediate case, with a companion mass that falls precisely between the two spider populations.

For each of the three systems, we obtain the first independent measurements of effective temperatures, measured solely from spectroscopy. We also present new radial velocity measurements for each system, and investigate the variety of spectral features from each companion and how they vary throughout the binary orbit.

\section{Observations, data, analysis}
\label{sec:obs+}

\subsection{Observations}
\label{sec:obs}

\begin{table*}
    \centering
    \caption{Log of GTC-OSIRIS long-slit spectroscopic observations.}
    \begin{tabular}{cccccccccc}
        \hline
        Object & $P_\mathrm{b}$ (h) & Exposures & Date & Grism & Wavelengths (\AA{}) & Phases & Airmass & Seeing (arcsec) & Slit (arcsec) \\
        \hline
        PSR J1048+2339 & 6.0 & $26 \times 915 \ \mathrm{s}$ & 2020-02-19 & R2500R & 5580 - 7680 & 0.6 - 1.7 & 1.0 - 1.9 & 0.8 - 2.0 & 1.0 \\
        PSR J1810+1744 & 3.6 & $22 \times 900 \ \mathrm{s}$ & 2015-06-28 & R1000B & 3630 - 7880 &  0 - 1.5  & 1.0 - 1.4 &  0.4 - 1.0 & 0.8 \\
        PSR J1908+2105 & 3.5 & $25 \times 900 \ \mathrm{s}$ & 2020-07-22 & R1000R & 5100 - 10400 &  0 - 1.8  & 1.0 - 1.6 & 0.4 - 0.8 & 1.0 \\
        \hline
    \end{tabular}
    \label{tab:obs}
\end{table*}

We obtained long-slit spectroscopic observations of all three targets using the Optical System for Imaging and low--intermediate Resolution Integrated Spectroscopy (OSIRIS) at the Gran Telescopio Canarias (GTC)\footnote{\url{http://www.gtc.iac.es/instruments/osiris/}}. Due to the short orbital periods of the targets, we were able to observe each system for at least one full orbit during a single night, in order to obtain complete phase coverage. Different grisms were selected for each target, in order to maximise the velocity resolution while maintaining an adequate signal-to-noise ratio (S/N) and a wavelength range covering a sufficient number of temperature-sensitive absorption line species. 

The chosen instrumental configurations for each target are given in Table~\ref{tab:obs}. The listed configurations resulted in resolutions of 3.0, 6.0, and 8.0\,\AA{}, or 140, 300, and 310\kms{} in velocity units, at the central wavelength of the observations for J1048, J1810, and J1908, respectively. The slit position angle was selected in order to include a nearby, brighter star in the slit with known magnitudes. For all observations, 2x2 binning was used to improve S/N. For J1048, the seeing degraded substantially towards the end of the observations, which had a significant effect on the precision of the final few spectra obtained. As such, the last three spectra were omitted from radial velocity and equivalent width measurements.

Standard bias-subtraction, flat-fielding, cosmic ray cleaning, and response corrections were carried out using the routines of the \textsc{starlink}\footnote{\url{http://starlink.eao.hawaii.edu/starlink/}} package \citep{starlink}. We performed optimal extraction and sky subtraction of 1D spectra from the 2D images using \textsc{starlink/pamela}\footnote{\url{http://cygnus.astro.warwick.ac.uk/phsaap/software/pamela/html/INDEX.html}} routines in order to correct for significant spectrum distortion \citep{marsh89}.

The wavelength calibration was performed using a combination of HgAr, Ne, and, for J1908 specifically, Xe calibration lamp frames, in order to precisely calibrate wavelengths across the full wavelength range. The positions of emission lines from the combined frames were measured and used to fit a polynomial to the dispersion relation. The resulting fits had rms residuals of 0.007, 0.03, and 0.05\,\AA{} with polynomials of order 9, 7, and 9 for J1048, J1810, and J1908 respectively.

With the exception of the observations of J1908, only one arc frame was taken, either at the beginning or end of the night. While GTC-OSIRIS has relatively high instrumental stability, in order to account for any possible drifts and further refine the wavelength solution, the positions of several bright telluric emission lines were traced throughout the observations. A weighted mean was taken across these lines to obtain the residual shifts to the wavelength solution, which were subsequently removed from the spectra. Throughout the observations, these shifts were on the order of 10\kms{}. 

For J1908, two arcs were taken, one before and one after the observations, and so a linear interpolation of the wavelength solution could be performed. A drift of $\sim$1\,\AA{} was found between the start and end arc frames. Subsequently, the same sky emission wavelength correction was applied to remove any additional drifts not accounted for by the linearly-interpolated solution.

\subsection{Radial velocity analysis}
\label{sec:rvs}

In order to precisely follow both the velocity and effective temperature of the companion stars in these systems, a robust orbital solution must first be determined. To do so, we applied cross-correlation techniques between the observed target spectra and a set of comparison, or `template' spectra. For the purpose of this analysis, a library of template spectra was built from the BT-Settl library of synthetic spectra \citep{allard11}, covering effective temperatures ranging from 2600 to 10000\,K. These synthetic spectra were normalised and broadened to match the instrumental resolution of each set of observations. 

We cross-validated the results obtained using these synthetic templates using a solar-metallicity subset of 34 stars from the UVES-POP library of VLT-UVES observations of main-sequence stars \citep{uves, linares18}, normalised and shifted to zero radial velocity. In all cases the synthetic templates were preferred, and provided better results, due to their even temperature coverage, constant metallicity, and constant surface gravity. As such, here we present results using only our BT-Settl template library. Further details on the template spectra used are given in Appendix~\ref{sec:templates}.

The cross-correlation was computed between the normalised target spectra and templates degraded to the same instrumental resolution, using the spectral analysis software \textsc{molly}\footnote{\url{http://cygnus.astro.warwick.ac.uk/phsaap/software/molly/html/INDEX.html}}. The initial search width used covered the radial velocity range of approximately $\pm 1000\kms{}$. In subsequent searches, once an orbital solution had been estimated, this was refined to $\pm 500\kms{}$ around the estimated radial velocity modulation. 
The wavelength ranges used for cross-correlation were chosen carefully, with spectral masks being constructed in order to measure radial velocities from various sets of absorption lines with different temperature sensitivities, and thus expected have different centres of light (see Section~\ref{sec:res} for details). 
These lines were identified primarily by the dependence of their equivalent widths on orbital phase, which can reveal where on the companion particular line features originate from. The ranges used for radial velocity and equivalent width measurements of each system are presented in Figures~\ref{fig:j1048}, \ref{fig:j1810}, and \ref{fig:j1908} for J1048, J1810, and J1908, respectively.

\begin{table}
    \centering
    \caption{Orbital ephemerides for each system. J1048 from \citet{deneva16}, J1810 from \citet{fermi23}, and J1908 from \citet{deneva21}.}
    \begin{tabular}{ccc}
        \hline
        Object & $P_\mathrm{b}$ (d) & $T_0$ (MJD) \\
        \hline
        PSR J1048+2339 & 0.250519045(5) & 56637.660807(1) \\ 
        PSR J1810+1744 & 0.1481702753(4) & 55130.109(8) \\ 
        PSR J1908+2105 & 0.1463168431(6) & 56478.320069(2) \\ 
        \hline
    \end{tabular}
    \label{tab:ephem}
\end{table}

Cross-correlation results were fit with a simple sinusoidal model for the radial velocity at time $t$: $v(t) = K_2 \sin{(2\pi (t-T_0)/P_\mathrm{b} )} + \gamma$, where $K_2$ is the orbital velocity semi-amplitude, $T_0$ is the time of companion inferior conjunction (where $\phi = 0$), and $\gamma$ is the systemic velocity. For all three systems, precise orbital ephemerides were already available \citep{deneva16, fermi23, deneva21} -- see Table~\ref{tab:ephem}. As such, the orbital period $P_\mathrm{b}$ was not fit, but instead kept fixed to the value from pulsar timing. However, due to the propagation of errors from the pulsar timing reference epochs to the observational epochs, we still fit for $T_0$, with the value from pulsar timing being used as an initial estimate. As such, the orbital model used has one fixed parameter, $P_\mathrm{b}$, and three free parameters, $K_2$, $\gamma$, and $T_0$.

\subsection{Optimal subtraction}
\label{sec:optsub}

In order to characterise the temperature variation of the companion stars throughout their orbits, we performed optimal subtraction \citep{marsh94} between the normalised target and template spectra (again, degraded to match the instrumental resolution). This process minimizes the residuals between the targets and templates, by optimising a `veiling factor' $f_\mathrm{veil}$ that scales the absorption lines of the template to account for contributions from non-stellar light. When performed across a library of template spectra covering a range of well-known stellar parameters, this allows for constraints on those parameters to be found. 

To determine companion temperatures, observations were first shifted in wavelength to remove orbital and systemic motion (as determined from cross-correlation measurements) and phase-binned to improve the S/N. In addition, we constructed spectral masks to focus the optimal subtraction on areas with line features corresponding to both the day and night sides of the companion, in order to better constrain temperatures. Further details on the phase bins and spectral masks used are given in Sections \ref{sec:j1048}, \ref{sec:j1810}, and \ref{sec:j1908}, for J1048, J1810, and J1908, respectively.

The result of performing optimal subtraction across a set of templates covering a range of effective temperatures provides values of $\chi^2$ as a function of \teff{}, for each phase-binned spectrum. In order to determine robust values of the effective temperature for each phase bin, we fit the $\chi^2$-minimum from each set of optimal subtraction data using the Python package \textsc{SciPy} \citep{scipy}. First, to identify points corresponding to the minimum, the optimal subtraction data points were approximated by a polynomial, to give a smoothly-varying function from which derivatives could be easily obtained. Then, the global minimum of this was identified from ascending zeroes of the first derivatives, and points were selected between the nearest ascending and descending zeroes of the second derivative. Finally, the selected points about the minimum were fit with a skewed normal distribution, which both provides a better fit to the minimum than the global polynomial approximation, and accounts for the asymmetry around the minimum which would not be captured by a Gaussian or quadratic fit. From this, we determine the best value of \teff{} along with 1-$\sigma$ errors, using a $\Delta\chi^2 = 1$ shift from the minimum.

\section{Results}
\label{sec:res}

\subsection{PSR J1048+2339}
\label{sec:j1048}

\begin{figure*}
    \centering
    \includegraphics[width=1\textwidth]{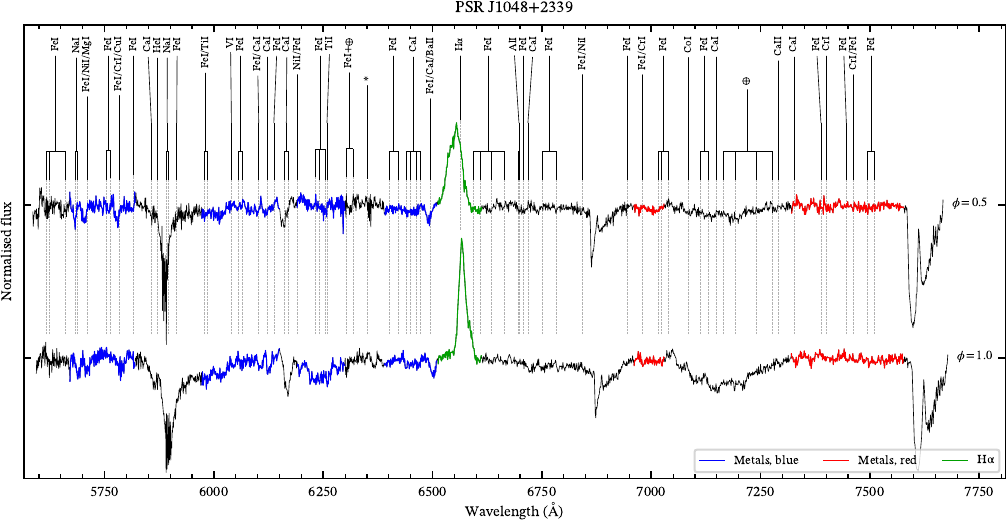}
    \caption{Normalised GTC-OSIRIS spectra of J1048 around companion superior conjunction (top) and companion inferior conjunction (bottom). The wavelength ranges used for equivalent width (Figure~\ref{fig:j1048ews}) and radial velocity (Figure~\ref{fig:j1048rvs}) measurements are highlighted, with colours as specified by the legend. The most prominent features have been identified and labelled along the top, based primarily on the Arcturus atlas of \citet{hinkle00}. The $\oplus$ symbol denotes absorption features primarily telluric in nature, while $\mathrm{\ast}$ indicates a defect in GTC-OSIRIS CCD2 centred at pixel Y$=765$.}
    \label{fig:j1048}
\end{figure*}

The optical spectra of J1048 presented in Figure~\ref{fig:j1048} are dominated by two main characteristics. The first is a substantial number of broad, blended absorption line species, most prominent of which are the Fe and Ca lines around 6200 and 6400\,\AA{} -- both typical of a K-dwarf. The second is a bright and broad H$\alpha$ emission line, which varies significantly in strength\footnote{We take here the standard convention where the strength of a line (in absorption or emission) is measured by its equivalent width, i.e., the width of the continuum with the same integrated flux as the line.} throughout the observed orbit, peaking at phases 0.1 and 0.5. We found no clear evidence of the \ion{He}{I} 6678\,\AA{} `accretion indicator' emission line in our GTC spectra, consistent with the 2020 VLT spectra presented by \citet{zanon21}. However, we did find signs of the \ion{He}{I} 5876\,\AA{} line in emission, as \citet{strader19} observed in some of their spectra from 2018, close to phase 0.1 and 0.5 -- seemingly coincident with the H$\alpha$ emission.

\begin{figure}
    \centering
    \includegraphics[width=1\columnwidth]{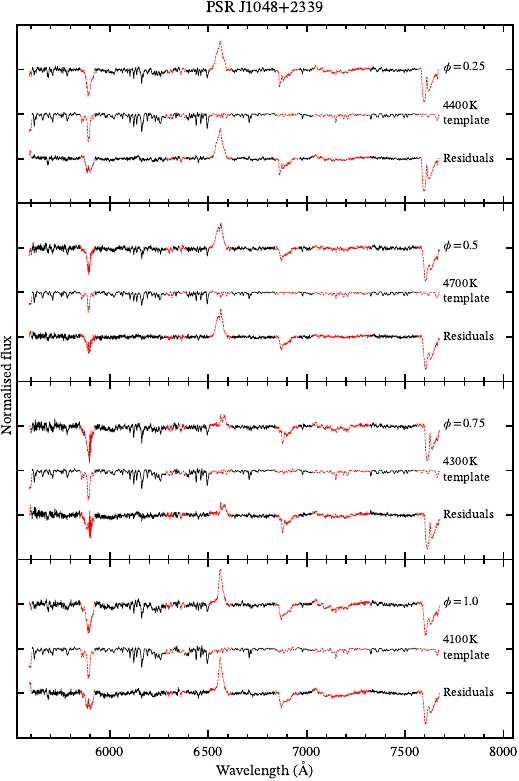}
    \caption{Optimal subtraction spectra for J1048 - best fit templates and residuals. From top to bottom: phases 0.25, 0.50, 0.75, and 1.0. Spectra were renormalised within these ranges to correct for the significant amount of blended lines (particularly in the 6200--6500\,\AA{} range) and so differ slightly in appearance to those presented in Figure~\ref{fig:j1048}. Excluded wavelength ranges are plotted in red.}
    \label{fig:j1048res}
\end{figure}

We performed optimal subtraction using the vast majority of the full wavelength range. The excluded ranges consist of the \ion{Na}{I} 5890+5896\,\AA{} doublet, which is contaminated by interstellar absorption, the broad H$\alpha$ emission line, a CCD blemish at approximately 6360\,\AA{}, and the various telluric absorption features throughout the spectral range. The extremities of the wavelength range were also masked to reduce noise introduced by end effects. The exact spectral range used for optimal subtraction is shown in Figure~\ref{fig:j1048res}.

Due to the substantial number of broadened and blended absorption features in the spectra of J1048, particularly around phase 0 and in the 6150--6400 and 6600--6850\,\AA{} ranges, an improved continuum normalisation was required in selected spectral regions before carrying out optimal subtraction. As such, the range selected for optimal subtraction was split into four regions, each of which was individually normalised with a low-order spline fit before being recombined into a full spectrum. We selected the edges of these regions to fall inside masked regions -- specifically the \ion{Na}{I} doublet, H$\alpha$ emission, and telluric absorption at 6900\,\AA{} -- so any discontinuities between each region would not affect optimal subtraction results. The same procedure was applied to the broadened template spectra to ensure consistency between the template and observational spectra. This `renormalisation' approach can allow for accurate optimal subtraction results for much cooler stars than is typically possible -- particularly for K- and M-dwarfs which often have wavelength ranges completely dominated by absorption, with no clear continuum.
We found this process resulted in a greatly improved optimal subtraction fit for J1048, which otherwise fails to match many of the absorption features, particularly around companion inferior conjunction. 

\begin{figure}
    \centering
    \includegraphics[width=1\columnwidth]{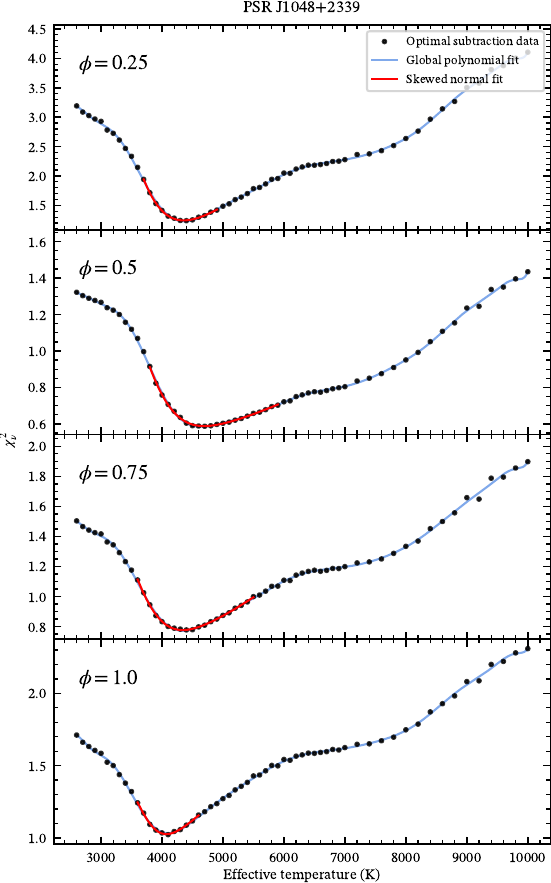}
    \caption{Optimal subtraction results for J1048 -- reduced chi-squared against effective temperature. From top to bottom: phases 0.25, 0.50, 0.75, and 1.0. Masked spectra and subtraction residuals are shown in Figure~\ref{fig:j1048res}. Red indicates the fit to the minimum used to determine effective temperatures and uncertainties for each phase bin.}
    \label{fig:j1048optsub}
\end{figure}

The results of optimal subtraction are shown in Figure~\ref{fig:j1048optsub}, with the fitting procedure outlined in Section~\ref{sec:optsub} applied. Observations of J1048 were combined and averaged together in four phase bins, each of a quarter phase in width. As such, four results were obtained: at companion superior conjunction ($\phi=0.5$), inferior conjunction ($\phi=1$), and both points of quadrature ($\phi=0.25$ and $\phi=0.75$). The phase-binned spectra along with their best-fitting templates and subtracted residuals are shown in Figure~\ref{fig:j1048res}. For the night side, we found a precise temperature of $T_\mathrm{night} = 4072^{+32}_{-31}$\,K, while we obtained a day side temperature of $T_\mathrm{day} = 4690^{+51}_{-48}$\,K. At quadrature, both measured temperatures are in excellent agreement: $T_\mathrm{q1} = 4356^{+33}_{-32}$\,K at $\phi=0.25$ and $T_\mathrm{q2} = 4339^{+42}_{-41}$\,K at $\phi=0.75$. Thus, J1048 varies in effective temperature by approximately 600\,K throughout its orbit, covering the spectral type range K3 to K7. All optimal subtraction results had a veiling factor $f_\mathrm{veil}$ of around 0.6, indicating about 40 per cent of the flux is non-stellar. These results mark the first spectroscopic temperatures measured from both the irradiated and non-irradiated sides of J1048, which we discuss further in Section~\ref{sec:temps}.

Radial velocities and equivalent widths were measured using two masks focused on different metallic line species. 
The first was selected from a specific range of absorption lines in the red end of the spectrum ($>6950$\,\AA{}), primarily consisting of Fe and Cr lines, which were found to be weakly correlated in equivalent width with temperature over the temperature range determined from optimal subtraction. 
The second mask was constructed from the wide range of absorption lines present at wavelengths shorter than H$\alpha$, which show a strong anticorrelation between equivalent width and temperature. 
In this range, the \ion{Fe}{I} lines around 6300\,\AA{}, the broad \ion{Ca}{I} line blend at 6160\,\AA{}, and the strong \ion{Na}{I} 5890+5896\,\AA{} doublet were excluded -- the first due to telluric contamination and a CCD defect, the second due to its unreliability in producing strong cross-correlations, and the third due to interstellar contamination. The relationship between equivalent width and effective temperature for the template spectra with these masks is shown in Figure~\ref{fig:temp_ews}. A third mask was also constructed, centred on the H$\alpha$ emission line, but was only used to measure its equivalent width variability throughout the orbit. The exact ranges of these masks are plotted in Figure~\ref{fig:j1048}. 

\begin{figure}
    \centering
    \includegraphics[width=1\columnwidth]{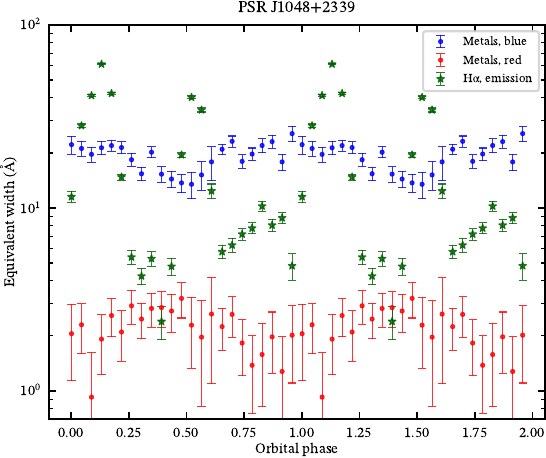}
    \caption{Equivalent widths of J1048 with orbital phase, from the red and blue metal and H$\alpha$ emission ranges shown in Figure~\ref{fig:j1048}. The sign of the H$\alpha$ equivalent widths has been inverted for display purposes. For the red metal mask, the range was refined to only include the identified lines in Figure~\ref{fig:j1048} to reduce noise. Three spectra from phases 0.7--0.8 have been omitted due to poor seeing, resulting in insufficient signal for reliable equivalent width measurements. Two orbits are plotted for clarity.}
    \label{fig:j1048ews}
\end{figure}

The equivalent widths measured with both the absorption and H$\alpha$ emission line masks are presented in Figure~\ref{fig:j1048ews}. The absorption lines captured by the red mask show a slight variation throughout the orbit, reaching a maximum near phase 0.5 and a minimum near phase 0. On the other hand, the second mask clearly follows the opposite trend, with a significant decrease in equivalent width around companion superior conjunction. This indicates that the line features captured in the blue mask are stronger towards the night side of the companion, while the red mask contains lines which are stronger towards the day side. 

An interesting feature present in both absorption line equivalent width curves is the asymmetry about their respective minima and maxima. For both masks, the extrema are centred approximately 10 per cent earlier in the orbit than expected, occurring closer to phases 0.4 and 0.9 than 0.5 and 1. This could be indicative of asymmetric heating or heat redistribution in the companion, as we discuss in more detail in Section~\ref{sec:temps}.

The equivalent widths from the H$\alpha$ emission line show two clear peaks: a large peak centred approximately at $\phi = 0.1$, and a smaller peak of shorter duration centred just after companion superior conjunction at $\phi = 0.5$. This trend is somewhat similar to that observed by \citet{zanon21}, although the emission line strength is much greater here. This emission line behaviour is discussed further in comparison to previous observations in Section~\ref{sec:misc}.

\begin{figure}
    \centering
    \includegraphics[width=1\columnwidth]{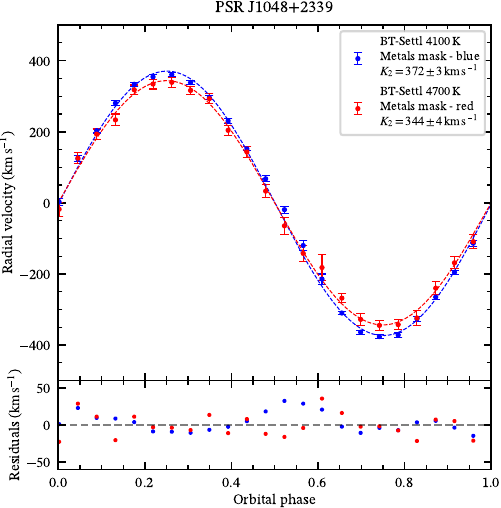}
    \caption{Radial velocity curves of J1048, from the red and blue metal ranges shown in Figure~\ref{fig:j1048}. The lower panel shows residuals to the sinusoidal fits. Three spectra from phases 0.7--0.8 have been omitted due to poor seeing, resulting in anomalous radial velocities that significantly affected the goodness of fit.}
    \label{fig:j1048rvs}
\end{figure}

The radial velocity curves obtained with both masks are presented in Figure~\ref{fig:j1048rvs}. Thanks to the abundance of strong, narrow absorption lines throughout the spectra, with the blue mask, we obtain a precise orbital velocity semi-amplitude of $K_\mathrm{metals, blue} = 372 \pm 3\kms{}$. With the red mask, we measure a lower value of $K_\mathrm{metals, red} = 344 \pm 4\kms{}$. 
This shows a clear and significant separation of the velocities measured from different absorption features, as was previously found in the redback system PSR J2215+5135 by \citet{linares18}.

For both fits, the systemic velocity and time of inferior conjunction are in good agreement with each other. Values from the fit with the red mask are $\gamma_\mathrm{metals, red} = 24 \pm 3\kms{}$ and $T_{0, \mathrm{metals, red}} = 58899.0960 \pm 0.0005$ MJD, while the blue mask fits have a slightly more precise$\gamma_\mathrm{metals, \ blue} = 23 \pm 2\kms{}$ and $T_{0, \mathrm{metals, blue}} = 58899.0963 \pm 0.0004$ MJD.

The sinusoidal fits give a $\chi^2_\nu$ of 0.82 for the fit with red mask and 2.8 for the fit with the blue mask, respectively. This indicates the errors on the blue mask model parameters are underestimated. As such, to obtain a more conservative estimate of the errors on the parameters for both fits, the radial velocity errors have been scaled to give a $\chi^2_\nu = 1$.

Both radial velocities curves were obtained by cross-correlating with templates representative of the day and night side of the companion, as measured by optimal subtraction. Thus, the blue- and red-mask radial velocities in Figure~\ref{fig:j1048rvs} were determined by cross-correlation against a 4100\,K and 4700\,K template, respectively. It is important to note that the same separation between the two masks is obtained regardless of the template spectra used.

\subsection{PSR J1810+1744}
\label{sec:j1810}

\begin{figure*}
    \centering
    \includegraphics[width=1\textwidth]{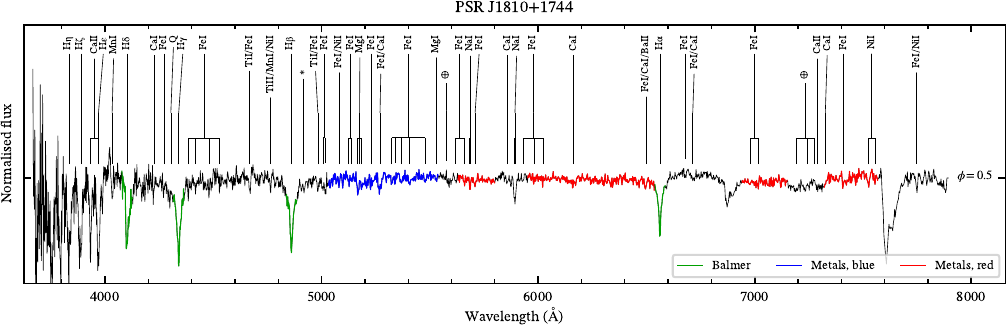}
    \caption{Normalised GTC-OSIRIS spectra of J1810 around companion superior conjunction, orbital phase = 0.5. The wavelength ranges used for equivalent width (Figure~\ref{fig:j1810ews}) and radial velocity (Figure~\ref{fig:j1810rvs}) measurements are highlighted, with colours as specified by the legend. The most prominent features have been identified and labelled along the top, based primarily on the Arcturus atlas of \citet{hinkle00}. The $\oplus$ symbol denotes absorption features primarily telluric in nature, while $\mathrm{\ast}$ indicates a defect in GTC-OSIRIS CCD2 centred at pixel Y$=765$.}
    \label{fig:j1810}
\end{figure*}

As J1810 is a heavily irradiated black widow, it shows substantial temperature variations throughout its 3.6-h orbit, with a stark contrast in appearance between its day and night sides. We observed J1810 over a broad spectral range, covering most of the optical wavelengths. This not only covers the entire Balmer series -- the most dominant feature in the spectra -- but it also captures a variety of metallic line features, such as the \ion{Mg}{I} triplet at 5167--5183\,\AA{}, alongside various Fe, Ca, Ni, and Na features at redder wavelengths, as shown in Figure~\ref{fig:j1810}.

\begin{figure}
    \centering
    \includegraphics[width=1\columnwidth]{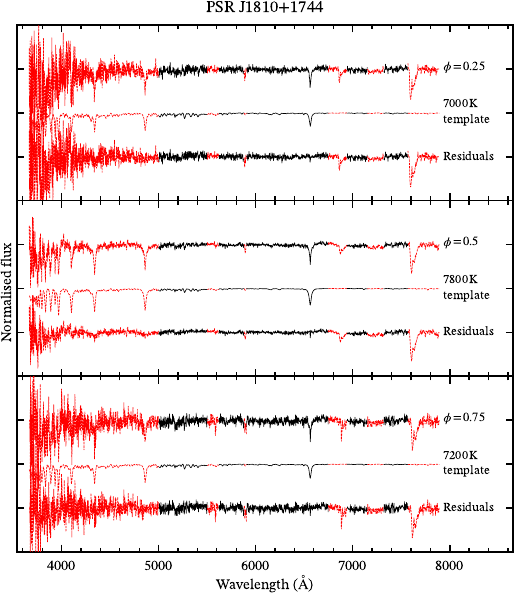}
    \caption{Optimal subtraction spectra for J1810 - best fit templates and residuals. From top to bottom: phases 0.25, 0.50, and 0.75. Excluded wavelength ranges are plotted in red.}
    \label{fig:j1810res}
\end{figure}

For optimal subtraction, the best results come from using the highest S/N regions of the spectra. As such, we excluded the blue end of the spectrum completely, where noise was too dominant ($<5000$\,\AA{}). In addition, the [\ion{O}{I}] 5577\,\AA{} line and \ion{Na}{I} 5890+5896\,\AA{} doublet were excluded to avoid contamination by telluric emission and interstellar absorption, respectively. The telluric molecular bands were also excluded. Thus, the optimal subtraction was focused on H$\alpha$, the \ion{Mg}{I} triplet, and various Fe, Ca, and Ni line species, all within the highest signal part of the spectra (see Figure~\ref{fig:j1810res}).

\begin{figure}
    \centering
    \includegraphics[width=1\columnwidth]{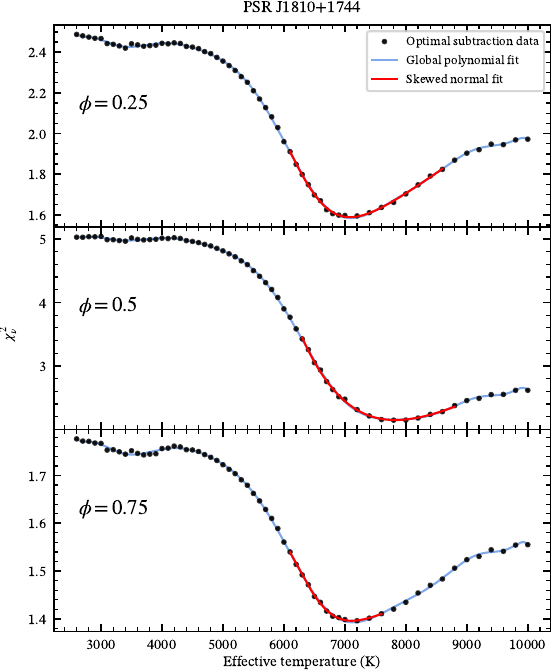}
    \caption{Optimal subtraction results for J1810 -- reduced chi-squared against effective temperature. From top to bottom: phases 0.25, 0.50, and 0.75. Phase 1.0 was omitted due to insufficient S/N for reliable measurements. Masked spectra and subtraction residuals are shown in Figure~\ref{fig:j1810res}. Red indicates the fit to the minimum used to determine effective temperatures and uncertainties for each phase bin.}
    \label{fig:j1810optsub}
\end{figure}

The spectra of J1810 were averaged together about phases 0, 0.25, 0.5, and 0.75, using phase bins of a quarter phase in width. The results of optimal subtraction performed on these phase-binned spectra are shown in Figure~\ref{fig:j1810optsub}, with the fitting procedure outlined in Section~\ref{sec:optsub} applied. At companion superior conjunction, we obtained a day-side temperature of $T_\mathrm{day} = 7827^{+90}_{-89}$\,K (spectral type A7). At both points of quadrature, the temperatures agree well: for $\phi=0.25$,  $T_\mathrm{q1} = 7085^{+87}_{-83}$\,K, and for $\phi=0.75$, $T_\mathrm{q2} = 7113^{+138}_{-125}$\,K (a spectral type of F2). The day-side fit had a veiling factor of $f_\mathrm{veil} = 0.87$, suggesting a small non-stellar contribution to the absorption line flux, while both points of quadrature had $f_\mathrm{veil} \simeq 1$. At companion inferior conjunction, the phase-binned spectrum corresponding to the dark side was dominated by noise, with no obvious absorption features throughout, so we could not measure reliable temperatures from the dark side. Nevertheless, these results still represent the first spectroscopic temperature measurements of J1810, both of its day side and at intermediate phases. The phase-binned spectra along with their best-fitting templates and subtracted residuals are shown in Figure~\ref{fig:j1810res}.

To measure radial velocities corresponding to each face of the companion, we constructed two masks for J1810 --  one sensitive to the hotter temperatures, consisting of 50\,\AA{} windows around the H$\alpha$, H$\beta$, H$\gamma$, and H$\delta$ lines, and one sensitive to colder temperatures, capturing the metallic absorption features in the red end of the spectra, distributed across the range 5600--7500\,\AA{}. 
Both masks are shown in Figure~\ref{fig:j1810}. 

\begin{figure}
    \centering
    \includegraphics[width=1\columnwidth]{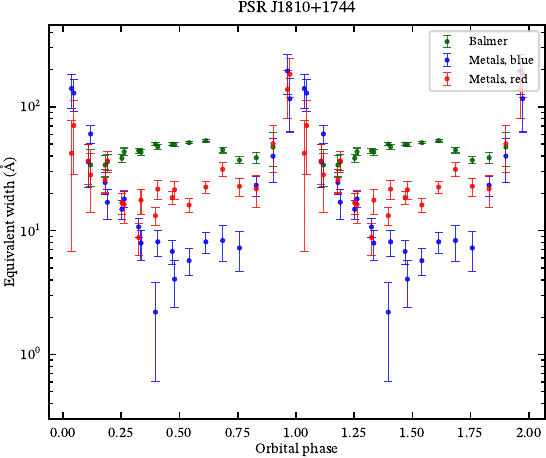}
    \caption{Equivalent widths of J1810 with orbital phase, from the Balmer, red, and blue metal ranges shown in Figure~\ref{fig:j1810}. Five points have been omitted from the Balmer equivalent widths around phase 0, where no significant measurement was found. A slight asymmetry is visible in all three series around phase 0.5. Two orbits are plotted for clarity.}
    \label{fig:j1810ews}
\end{figure}

Measuring equivalent width as a function of orbital phase for each mask reveals their variability over the surface of the companion and their sensitivity to temperature. 
As shown in Figure~\ref{fig:j1810ews}, the Balmer lines are strongest about companion superior conjunction, when the irradiated face is visible, and vanish entirely at phase 0. Metallic features are strongest around companion inferior conjunction, and weaken significantly around phase 0.5. 
Again, a slight asymmetry is seen around phase 0.5: the maximum of the hydrogen lines is centred just before phase 0.5. This equivalent width asymmetry is further discussed in Section~\ref{sec:temps}.

Additionally, we considered a third mask covering metallic absorption features at bluer wavelengths, in the range 5020--5540\,\AA{} (see Figure~\ref{fig:j1810}) -- its equivalent width vs. orbital phase relationship is also shown in Figure~\ref{fig:j1810ews}. This mask is evidently even more sensitive to colder temperatures, changing much more sharply with orbital phase than the other metallic mask, suggesting the \ion{Mg}{I} triplet is a prominent species of the non-irradiated surface of the companion. However, because of the significantly reduced flux at the blue end of the spectrum, due to both the low temperature and  the faintness of the dark side, and the reduced grism and detector efficiencies at short wavelengths, this mask did not prove suitable for measuring precise radial velocities for J1810.

\begin{figure}
    \centering
    \includegraphics[width=1\columnwidth]{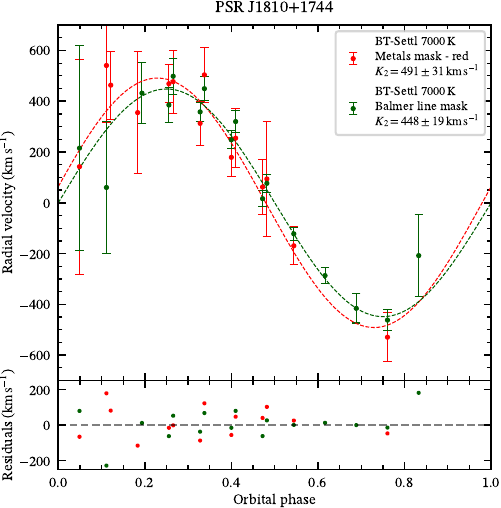}
    \caption{Radial velocity curves of J1810 from the Balmer and red metal ranges shown in Figure~\ref{fig:j1810}. Points with no significant cross-correlation signal have been excluded. The lower panel shows residuals to the sinusoidal fits.}
    \label{fig:j1810rvs}
\end{figure}

The radial velocity curves obtained with the two masks are shown in Figure~\ref{fig:j1810rvs}. With the Balmer mask, we determined an orbital velocity semi-amplitude of $K_\mathrm{Balmer} = 448 \pm 19\kms{} $, whereas a higher, but less precise $K_\mathrm{metals} = 491 \pm 32\kms{} $ was measured with the metal line mask. While the measured difference between $K_\mathrm{Balmer}$ and $K_\mathrm{metals}$ is not large enough to be significant, the fact that $K_\mathrm{Balmer} < K_\mathrm{metals}$ is consistent with Balmer and metallic lines dominating the hot and cold sides of the companion, respectively. The precision of the orbital velocity solution is mainly limited by the substantial reduction in flux around companion inferior conjunction -- very few good cross-correlations could be found close to phase 0, as is seen in Figure~\ref{fig:j1810rvs}.

The values of $T_0$ and $\gamma$ for the two fits are also consistent with each other within their uncertainties. For the fit with the Balmer mask, we obtained $\gamma_\mathrm{Balmer} = -142 \pm 26\kms{}$ and $T_{0, \mathrm{Balmer}} = 57202.097 \pm 0.002$ MJD, while for the metallic line mask fit we found $\gamma_\mathrm{metals} = -160 \pm 30\kms{}$ and $T_{0, \mathrm{metals}} = 57202.094 \pm 0.002$ MJD.

As might be expected, the best orbital solution for the Balmer mask was obtained when cross-correlating against a high temperature template with strong hydrogen absorption features -- the result using a $\teff{} = 7000$\,K template is shown, corresponding to an F2 spectral type. This result with a $7000$\,K template was selected based on being the closest match to the optimal subtraction result at quadrature, and its excellent goodness-of-fit, with $\chi^2_\nu = 1.16$.  This template is well-representative of the companion for a majority of its orbit, and thus was used for the metal line fit also. This resulted in a fit with $\chi^2_\nu = 0.45$, indicating the errors from cross-correlation are likely over-estimated. As such, to obtain a more conservative estimate of the errors on the fit parameters, the radial velocity errors have been suitably scaled to give a $\chi^2_\nu = 1$.

\subsection{PSR J1908+2105}
\label{sec:j1908}

\begin{figure*}
    \centering
    \includegraphics[width=1\textwidth]{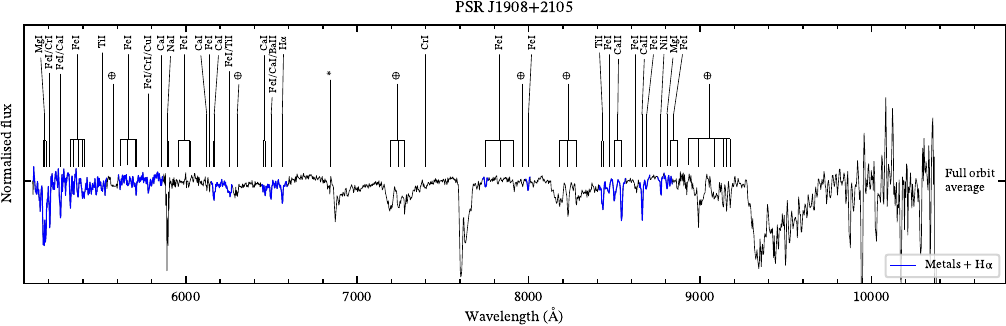}
    \caption{Normalised GTC-OSIRIS spectra of J1908, averaged across one full orbit. The wavelength ranges used for equivalent width (Figure~\ref{fig:j1908ews}) and radial velocity (Figure~\ref{fig:j1908rvs}) measurements are highlighted. The most prominent features have been identified and labelled along the top, based primarily on the Arcturus atlas of \citet{hinkle00}. The $\oplus$ symbol denotes absorption features primarily telluric in nature, while $\mathrm{\ast}$ indicates a defect in GTC-OSIRIS CCD2 centred at pixel Y$=765$.}
    \label{fig:j1908}
\end{figure*}

The optical (and near-infrared) spectra of J1908, shown in Figure~\ref{fig:j1908}, have the appearance of a relatively typical late G-type main sequence star. H$\alpha$ is clearly visible as one of the more prominent absorption features in the optical range, but is not broad nor particularly strong -- comparable in strength to the blended Fe/Ca/Ba metal line at 6496\,\AA{}. Most pronounced are the Mg triplet and nearby Fe lines, located at the bluest end of the spectra -- although the former suffers from substantial vignetting due to being located right at the edge of wavelength range. In the near-infrared end of the spectrum however, between the many telluric absorption bands, the \ion{Ca}{II} triplet is clearly observed at 8498, 8542, and 8662\,\AA{} as the three most well-resolved features, between a small variety of weaker metal lines. 

\begin{figure}
    \centering
    \includegraphics[width=1\columnwidth]{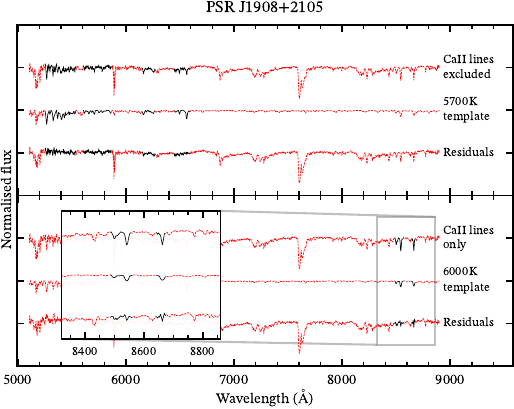}
    \caption{Optimal subtraction spectra for J1908 - best fit templates and residuals. The top and bottom panel show the results including and omitting, respectively, the \ion{Ca}{II} lines identified at 8498, 8542, and 8661\,\AA{} in Figure~\ref{fig:j1908}. An inset plot is included in the bottom panel, which shows a magnified view of the \ion{Ca}{II} lines. Excluded wavelength ranges are plotted in red.}
    \label{fig:j1908res}
\end{figure}

We determined line-based temperatures for J1908 from the most prominent absorption lines identified in Figure~\ref{fig:j1908}. The \ion{Mg}{I} triplet, seen at the far blue edge of the spectra, was the only exception to this, and was excluded because of significant uncertainties from heavy vignetting at the edge of the wavelength range. We fit the near-infrared \ion{Ca}{II} triplet separately from the optical lines, as we found the deep \ion{Ca}{II} lines could not be reproduced while simultaneously fitting the optical absorption lines. Thus, optimal subtraction was performed twice for J1908, once with the optical metallic and H$\alpha$ absorption lines, and again only considering the \ion{Ca}{II} triplet (see Figure~\ref{fig:j1908res}).

\begin{figure}
    \centering
    \includegraphics[width=1\columnwidth]{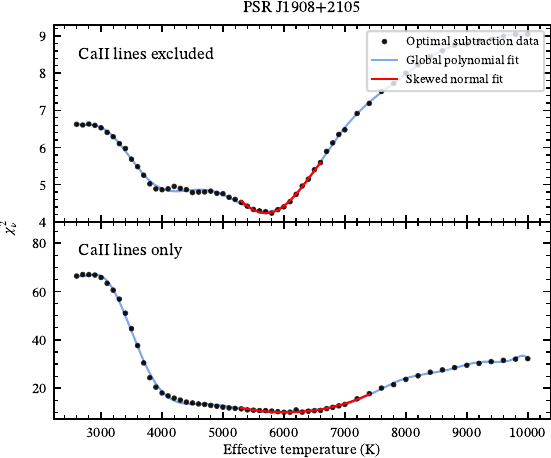}
    \caption{Optimal subtraction results for J1908 -- reduced chi-squared against effective temperature. The top and bottom panel show the results including and omitting, respectively, the \ion{Ca}{II} lines identified at 8498, 8542, and 8661\,\AA{} in Figure~\ref{fig:j1908}. Masked spectra and subtraction residuals are shown in Figure~\ref{fig:j1908res}. Red indicates the fit to the minimum used to determine effective temperatures and uncertainties for each phase bin.}
    \label{fig:j1908optsub}
\end{figure}

The optimal subtraction results for J1908 are presented in Figure~\ref{fig:j1908optsub}. Unlike with J1048 and J1810, we found no significant temperature variations when performing optimal subtraction on phase-binned spectra, with a measured $T_\mathrm{day} = 5638^{+97}_{-96}$\,K for the day side and $T_\mathrm{night} = 5627^{+93}_{-92}$\,K for the night side. Because of this, all spectra were averaged together across the entire orbit to produce a full-orbit averaged spectrum of J1908. Using this full-orbit average, we determined the effective temperature to be $\teff{} = 5706^{+72}_{-73}$\,K (spectral type G4) when excluding the \ion{Ca}{II} lines. This is the first spectroscopic temperature measurement of the optical counterpart of this system. A much less precise, but still consistent fit was obtained from the \ion{Ca}{II} lines of $\teff{} = 6042^{+291}_{-294}$\,K. Of much greater interest is the veiling factor, $f_\mathrm{veil}$, obtained from the two fits. For the first mask, we found a fairly typical $f_\mathrm{veil}$ of $0.67\pm0.01$, suggesting an approximately 30 per cent contribution from non-stellar light. However, for the \ion{Ca}{II} triplet, we obtained a substantially higher $f_\mathrm{veil}$ of $1.65\pm0.03$, indicating that \ion{Ca}{II} lines are enhanced compared to the template absorption lines. Possible mechanisms for producing such an enhancement are discussed in Section~\ref{sec:misc}.

We found the radial velocity and equivalent width variability to be very small throughout the orbit of J1908. While attempts were made measure separate radial velocities from the different line species of the companion, no significant differences were found between the variability of the various absorption features present in the spectra. This could be indicative of a system with little to no irradiation, or a very low orbital inclination. As such, only one mask was considered for measuring equivalent widths and radial velocities. This mask is shown in Figure~\ref{fig:j1908}, and covers all of the significant absorption features throughout the spectral range. 

\begin{figure}
    \centering
    \includegraphics[width=1\columnwidth]{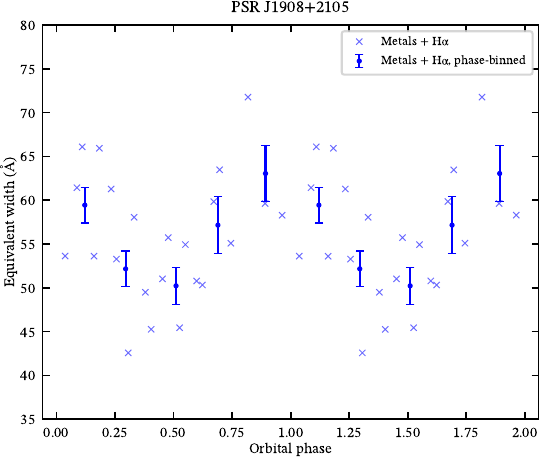}
    \caption{Equivalent widths of J1908 with orbital phase, from the wavelength ranges highlighted in Figure~\ref{fig:j1908}. Equivalent widths have been phase-binned into five bins in order to increase the prevalence of any phase-dependence. Two orbits are plotted for clarity.}
    \label{fig:j1908ews}
\end{figure}

The equivalent widths measured with this mask are plotted in Figure~\ref{fig:j1908ews}. The equivalent width variations are very small throughout the orbit, at the level of 10--15 per cent. To improve the visibility of these variations, we binned the individual measurements into five phase bins, 20 per cent of an orbit in width. Subsequently, this reveals a subtle trend in the absorption features captured by the metals + H$\alpha$ mask, which peak at companion inferior conjunction. The equivalent widths trace a smooth, sinusoidal shape, modulated at the orbital period from radio timing \citep{deneva21}, as might be expected for a low-inclination system, with no observable sharp transitions between the inner and outer faces of the companion. 

\begin{figure}
    \centering
    \includegraphics[width=1\columnwidth]{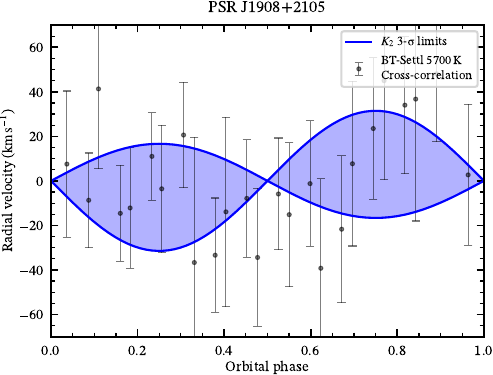}
    \caption{Radial velocities from J1908, measured from the wavelength ranges highlighted in Figure~\ref{fig:j1908}. 3-$\sigma$ upper limits on the semi-amplitude of the radial velocity curve, $K_2$, are highlighted in blue.}
    \label{fig:j1908rvs}
\end{figure}

The radial velocity measurements for J1908 are plotted in Figure~\ref{fig:j1908rvs}, obtained by cross-correlating against a 5700\,K template. We performed multiple sinusoidal fits on the radial velocities: with $T_0$ as a free parameter, with $T_0$ fixed to the radio timing value given in Table~\ref{tab:ephem} from \citet{deneva21}, and also before and after using the telluric modelling code \textsc{TelFit}\footnote{\url{http://telfit.readthedocs.io/}} to remove any residual telluric motion. We found no significant sinusoidal signal, when using the 5700\,K template or with any other template spectra.

Instead, a 3-$\sigma$ upper limit on $K_2$ was determined, as highlighted in blue in Figure~\ref{fig:j1908rvs}. This was found by fixing $\gamma$ to $-41\kms{}$, from the weighted mean of the radial velocity points, and calculating $\chi^2$ as a function of $K_2$ over a $\Delta \chi^2$ of 9 from the minimum (corresponding to a 3-$\sigma$ confidence interval for $\nu=1$). With this we determine an upper limit of $K_2 < 32\kms{}$, lower than that of any other known or candidate spider system. We discuss the implications of this for the dynamical properties of J1908 in Section~\ref{sec:masses}.

\section{Discussion}
\label{sec:disc}

\subsection{Velocity measurements and $K$-correction}
\label{sec:vels}

The precise measurement of the radial velocity semi-amplitude $K_2$ is paramount to obtaining robust mass constraints on the pulsars in spider systems. Unfortunately, radial velocities measured from Doppler-shifted absorption lines in irradiated systems contain a crucial systematic error that must be accounted for: the difference between the centre of mass (CoM) and centre of light of the companion. This effect arises due to the uneven surface temperature of the companion, causing some absorption lines on the irradiated face to be enhanced, while others are `quenched' and thus become more biased towards the cold side. Those absorption features that are enhanced, and become stronger on the irradiated day side, will give lower radial velocities when measured, while those that are stronger towards the night side will give higher radial velocities. As such, this must be accounted for in order to obtain accurate masses, particularly in systems with heavy irradiation and compact orbits.

Traditionally, this has been handled by applying the $K$-correction, adjusting the observed $K_2$, $K_{2, \mathrm{obs}}$ by a factor $K_{2, \mathrm{CoM}}/K_{2, \mathrm{obs}}$ to obtain the actual, dynamical $K_2$, $K_{2, \mathrm{CoM}}$ \citep{wade88}. The correction factor needed is typically calculated from models, or estimated in systems where irradiation is a transient phenomenon. However, such solutions can carry substantial errors with them, which are then propagated into the mass constraints. Instead, here we adopt the `empirical $K$-correction' introduced by \citet{linares18}, whereby absorption features from the day and night side are used to `bracket' the true value of $K_{2, \mathrm{CoM}}$. 

We note that this picture does not necessarily apply to line \textit{fluxes}, as a higher day-side continuum flux can pull the centre of light of low-temperature species away from the night-side \citep{dodge24}. In a lightly-irradiated system (such as J1048), however, this effect is unlikely to be so large that the centre-of-mass $K_2$ is no longer bracketed. In addition, in heavily-irradiated systems (e.g. J1810 and PSR J2215+5135), the absorption features from the colder parts of the companion will still have a centre of light close to the centre of mass \citep[see][Figure 5]{dodge24}. For J1810, such subtleties are absorbed by our statistical uncertainties. A full, self-consistent treatment of these effects requires careful, combined (and computationally expensive) photometric and spectroscopic modelling, beyond the scope of the present work. In light of this, we note these potential caveats and proceed by adopting the `empirical $K$-correction' for our systems.

Considering first the radial velocity curves measured for J1048, the two radial velocity curve semi-amplitudes of $K_\mathrm{metals, red} = 344 \pm 4\kms{}$ and $K_\mathrm{metals, blue} = 372 \pm 3\kms{}$ show a clear separation of the high and low temperature absorption features, which we therefore expect to bracket the centre-of-mass $K_2$ well. 
Both of our measured semi-amplitudes agree well with previous observations of J1048:

\citet{strader19} measured a $K_2$ of $376 \pm 14\kms{}$ from SOAR-Goodman spectra taken in 2018, in excellent agreement with our $K_\mathrm{metals, blue}$. They do not specify the exact wavelength ranges used for cross-correlation, only that it is relatively wide and includes many of the broad absorption features present in the optical spectra. In our spectra, we see that most of the dominant absorption lines are strongest around the night side. We suspect the same is true for their observations, and thus the radial velocities they measure are biased towards the night side -- however, without knowing the exact wavelengths used it is impossible to draw any substantial conclusions from their data. 

\citet{zanon21} found a lower $K_2$ of $343.3 \pm 4.4\kms{}$, almost identical to the $K_\mathrm{metals, red}$ we measured here. Their radial velocities were measured from two narrow wavelength ranges: from 5970 to 6291\,\AA{}, and between 6421 and 6522\,\AA{}. Surprisingly, these wavelength ranges are very similar to those included in our blue metals mask (see Figure~\ref{fig:j1048}), which we used to measure radial velocities from the night side. The two main differences are that we included some bluer metallic lines around 5750\,\AA{}, and we excluded the blended \ion{Ca}{I} lines at 6160\,\AA{} --  we found the latter to be unreliable as the blended components vary differently in strength with temperature. Therefore, we suspect the inclusion of this blended feature biases their measured radial velocities towards the day side of the companion, while the bluer metallic features we include have the opposite effect.

The radial velocities obtained for J1810 show $K_2$ in the range 448--490\kms{}, as well as a qualitatively similar separation to that found in J1048, with a lower $K_2$ measured from absorption features associated with higher temperatures. Although the difference is not large enough to be significant -- due to the substantial errors on the fit parameters -- it is still consistent with Balmer lines from the hot, irradiated inner face and metallic lines from the colder regions of the companion's surface. Comparing these values with that reported by \citet{romani21} from photometric modelling with radial velocity curves used for marginalisation: $K_\mathrm{2, CoM} = 462.9 \pm 2.2\kms{}$, it is clear that our observed values agree well with their results. However, it is also clear that the measured orbital velocity of J1810 could be greatly improved with deeper observations of the dark side. Observing in the near-infrared and focusing on features such as the strong \ion{Ca}{II} triplet seen in J1908's spectra could be more fruitful, particularly for black widows where most of the companion's night side flux is emitted at longer wavelengths.

As outlined in Section~\ref{sec:j1908}, we found no significant radial velocity variability from the optical counterpart of J1908. Instead, we determined a 3-$\sigma$ upper limit of $K_2 < 32\kms{}$ by varying $K_2$ and fixing $\gamma$, $T_0$, and $P_\mathrm{b}$. Thus, this upper limit assumes the radio ephemeris from \citet{deneva21} is still accurate for J1908 at the epoch of the optical observations. A simple error propagation suggests this to be true, giving an error on the orbital phase of $\sigma_\phi = 7 \times 10^{-5}$. 

We also consider the possibility of orbital period variations in J1908, as found in J1048 \citep{deneva16}. If $T_0$ is also a free parameter in determining the upper limit for $K_2$, this raises the degrees of freedom to $\nu = 2$, and the obtained upper limit becomes $K_2 < 49\kms{}$. However, the radio observations of J1908 cover a time span of 11 years, and it would appear that no orbital period derivatives were required to fit the data \citep{deneva21}. Additionally, the equivalent width variations in Figure~\ref{fig:j1908ews} would also suggest the orbit is not out of phase with respect to this solution. As such, the assumption of a fixed $T_0$ should hold true for J1908, and the upper limit of $K_2 < 32\kms{}$ is considered to be robust.

\subsection{Temperature variations}
\label{sec:temps}

Spider pulsars show a wide range of irradiation characteristics, and thus temperature variations, throughout the population. While about half of the known redbacks and nearly all known black widows are strongly irradiated, some systems show little or no evidence of irradiation \citep{turchetta23}. We have shown that leveraging the myriad of absorption lines in spider spectra, particularly with optimal subtraction techniques, can provide excellent temperature sensitivity when paired with a set of well-characterised template spectra.

J1048 is a particularly interesting case, being known to transition between an irradiated and non-irradiated state in less than 14 days \citep{yap19}. For J1048, simultaneous photometry clearly shows the system to be in its irradiated state \citep{tidemann23}, and indeed this is confirmed by the observed difference between $T_\mathrm{day}$ ($= 4690^{+51}_{-48}$\,K) and $T_\mathrm{night}$ ($= 4072^{+32}_{-31}$\,K) of approximately 600\,K. The temperatures we measured correspond to a spectral type of K3 at superior conjunction and K7 at inferior conjunction, agreeing well with the full-orbit average determined by \citet{zanon21} of K8$\pm$2, the base temperature from \citet{yap19}, and the day-side temperature determined from simultaneous photometry ($\sim$K2). As we noted in Section~\ref{sec:vels}, the dominant features in the optical spectra of J1048 are the metallic emission lines from the night side. Therefore, the full-orbit average temperature from \citet{zanon21}, determined from optimal subtraction (which relies on matching absorption lines), is likely biased towards the night side -- and thus agrees best with our $T_\mathrm{night}$. 

Interestingly, \citet{zanon21} do not find any significant temperature difference between the day- and night-side temperatures. They do, however, observe an equivalent width modulation that would suggest temperature variations throughout the orbit. It is unclear if they considered phase-binned spectra in their analysis, or if their conclusion comes only from the individual spectra -- if the latter is true, the S/N may have been insufficient to measure the day- and night-side temperature difference. This could also be due to a change in irradiation between the observation epochs -- although the single-peaked asymmetric light curve with $\Delta R \sim 0.9$\,mag, identical to that of our simultaneous photometry \citep{tidemann23}, would appear to indicate otherwise. Repeating this analysis while the system is in a non-irradiated state could prove insightful, particularly with regards to measuring the centre-of-light shift of absorption lines due to irradiation.

J1810, as is the case for the vast majority of black widows, is heavily irradiated. Despite being a low-mass star, we observe temperatures of $T_\mathrm{day} = 7827^{+90}_{-89}$\,K from its irradiated face, with a spectrum dominated by hydrogen absorption lines. While the S/N was too low for a reliable temperature measurement of the night side, it is clear from the large decrease in signal and the strong rise in metallic line equivalent widths (see Figure~\ref{fig:j1810ews}) that the temperature variation is substantial for J1810. This would appear to support the photometric modelling of \citet{romani21}, which finds a base temperature for the companion of around 3500\,K. However, direct spectroscopic constraints on the night-side temperature and velocity of J1810 are still desirable, and would lead to a more robust model solution. Until then, in the absence of well-measured velocities and temperatures of the dark side, the masses determined by \citet{romani21} remain uncertain (i.e. susceptible to systematic errors) and must be taken with caution.

On the other hand, we did not observe any significant temperature variation from the spectra of J1908 throughout its orbit. There are two main possibilities for this. The first is that J1908 is simply a system with negligible irradiation. However, the sinusoidal equivalent width modulation we found gives evidence for temperature changes across the companion surface (see Section~\ref{sec:j1908}, Figure~\ref{fig:j1908ews}). The second explanation is that J1908 is seen at low inclination, with a near face-on orbit. Strong irradiation is still possible, but since the observed spectrum is an average in flux over the visible hemisphere, the effect would become much more subtle. Considering the lack of a significant radial velocity signal measured from J1908, and the single-peaked low-amplitude light curves from \citet{beronya23}, the second hypothesis seems more likely.

From the full-orbit average spectrum, we measured a temperature of $\teff{} = 5706^{+72}_{-73}$\,K, corresponding to a spectral type of G4. This is significantly higher than the values \citet{beronya23} report from photometry ($4600\pm250$\,K at 2\,kpc, $4900\pm300$\,K at 5\,kpc) and spectroscopy (spectral type K--M). However, \citet{beronya23} only use a visual comparison with K5 and M1 spectral templates, and their photometric temperatures change substantially with the assumed distance, extinction, and Roche-lobe filling factor. Optimal subtraction relies on matching the depths of absorption lines, and so is not subject to these assumptions. As such, we consider our independent line-based temperature for J1908 to be more robust.

Equivalent width measurements not only prove a useful tool for discerning the temperature sensitivity (and hence phase dependence) of various line species, but also reveal some surprising effects. Both J1048 and J1810 show evidence of asymmetries in their equivalent width curves and, by extension, their heating. 

In J1810, the Balmer line maximum peaks after phase 0.5, with a gradual increase up to the maximum followed by a sharper decline. Similar asymmetries have been observed before: in the light curves of \citet{romani21}, the maximum follows a similar shape to that observed here in the equivalent widths, and also in the radio eclipses observed by \citet{polzin18}, which show a significant asymmetry suggestive of a swept tail of material. While it is challenging to favour any specific mechanism for creating these asymmetries based on the equivalent width data alone, one possible explanation would be that the intrabinary shock is wrapped around the leading side of the companion, resulting in asymmetric heating of the day side, which is then convected across the face. This would lend support to the models of \citet{romani21}, which favour a wind to redistribute heat across the surface of the companion.

On the other hand, in J1048 both sets of absorption features have maxima centred earlier in phase, ahead of phase 0.5 and 1, respectively. The equivalent width minima are also shifted earlier in phase. This is consistent with the asymmetric light curve from the simultaneous photometry of \citet{tidemann23}. 
Such an asymmetry might be expected if the intrabinary shock, which X-ray light curves suggest is wrapped around the pulsar \citep{cho18}, sweeps slightly retrograde in the orbit and thus heats the companion's trailing edge more. A shock geometry like this could be produced if the companion's wind dominates over the pulsar's \citep{romani16}, which could be the case in J1048 due to the relatively low spin-down luminosity of the pulsar: $\dot E \simeq 1 \times 10^{34}\,\mathrm{erg \ s^{-1}}$ \citep{deneva16}.

It is clear from these results that equivalent widths provide a strong proxy of the companion's temperature -- in all three cases, they modulate in nearly identical ways to the broad band photometry of their systems. For both J1048 and J1810, this extends to the asymmetries present in their photometry, and demonstrates equivalent width measurements are an effective way to trace temperature variations at a better time resolution than optimal subtraction of phase-averaged spectra can provide. 

\subsection{Importance of line-based temperatures}
\label{sec:linetemps}

Line-based temperatures of spiders are crucial for determining precise neutron star masses. While photometric modelling of spider light curves can produce good results, significant degeneracies exist in the fitting process. Companion effective temperature, extinction, distance, and inclination are all difficult to disentangle when fitting light curves. Line-based temperature measurements can provide independent constraints on the companion's effective temperature which can help lift these degeneracies, resulting in a robust inclination measurement and thus giving a robust neutron star mass.

When applying line-based temperature constraints, it is important to consider that the observed temperatures themselves \textit{also} depend on inclination, as well as the integration time of the observations and any phase-binning used to improve S/N. Indeed, the temperature we derive from any given spectrum can be thought of as a flux-weighted average over the visible elements of the star (which can be of vastly different temperatures), averaged again over the length of the exposure.
The effective temperature of the companion measured at quadrature can offer a different angle for constraining the properties of the companion: in the absence of asymmetric heating effects, i.e. a direct heating model, it would be inclination independent.  
This is due to exactly half of both the inner and outer faces being visible at phases 0.25 and 0.75.
Thus, it could allow the companion's intrinsic temperature and irradiation luminosity to be constrained independently of its inclination.
It is clear, however, that the assumption of isotropic heating is not valid in many systems. In such cases, averaging the temperatures at both points of quadrature ($\left < T_\mathrm{q1}, T_\mathrm{q2} \right > = T_\mathrm{q}$) could counteract this, again giving an inclination-independent temperature even where asymmetries are present.

To test this, we used \textsc{PHOEBE} v2.4.14\footnote{\url{http://phoebe-project.org}} \citep{phoebe2, phoebe2.3} to simulate observations of spider systems at varying inclinations. Three systems were constructed, taken as representative of the general properties of the spider population: a weakly-irradiated redback, a strongly-irradiated redback, and a strongly-irradiated black widow. For all systems, the neutron star mass was set to $1.8\ms{}$. The PHOENIX model atmospheres \citep{phoenix} were used for the companion and a perfect albedo of 1 for incident irradiation was assumed. For the two strongly-irradiated systems, the effect of asymmetric heating was also considered by introducing a hotspot, situated along the companion's equator, $45^\circ$ from its nose, towards the leading face. This hotspot had a radius of $25^\circ$ and was heated to 1.2 times the local surface temperature. Thus, 5 configurations were tested in total, with input parameters given in Table~\ref{tab:params}.

\begin{table}
    \centering
    \caption{Input parameters for the 5 model systems considered. $T_\mathrm{base}$ and $T_\mathrm{irr}$ are the base and irradiation temperatures of the companion, respectively. $A_\mathrm{HS}$ is temperature of the hotspot relative to its local intrinsic temperature, while $r_\mathrm{HS}$ is its radius. $\theta_\mathrm{HS}$ and $\phi_\mathrm{HS}$ specify its location as its colatitude and longitude respectively.}
    \begin{tabular}{lccccc}
        \hline
        Parameters & RB, low irr & RB & RB, asym & BW & BW, asym \\
        \hline
        $q$ & 0.2 & 0.2 & 0.2 & 0.02 & 0.02 \\
        $T_\mathrm{base} \ (\mathrm{K})$ & 5000 & 5000 & 5000 & 3000 & 3000 \\
        $T_\mathrm{irr} \ (\mathrm{K})$ & 3000 & 8000 & 8000 & 8000 & 8000 \\
        $P_\mathrm{orb} \ (\mathrm{hr})$ & 6.0 & 6.0 & 6.0 & 4.0 & 4.0 \\
        $A_\mathrm{HS}$ & - & - & 1.2 & - & 1.2 \\
        $r_\mathrm{HS} \ (^\circ)$ & - & - & 25 & - & 25 \\
        $\theta_\mathrm{HS} \ (^\circ)$ & - & - & 90 & - & 90 \\
        $\phi_\mathrm{HS} \ (^\circ)$ & - & - & -45 & - & -45 \\
        \hline
    \end{tabular}
    \label{tab:params}
\end{table}

For each of the 5 configurations, the inclination $i$ was varied from 0$^\circ$ to 90$^\circ$ in 2$^\circ$ increments. Also, we tested four different values for the Roche-lobe filling factor ($=x_\mathrm{nose} / x_\mathrm{L1}$): 0.7, 0.8, 0.9, and 0.99. 
For each system in the model grid, we computed the companion's surface temperature at specified points in its orbit. To simulate observations, the temperatures at several points were averaged together, weighted by flux, in phase bins of a quarter phase in width and centred around phases 0, 0.25, 0.5, and 0.75. We determined that averaging the companion's temperature over 100 points per phase bin approximated a continuous observation well, as increasing the number beyond this had no appreciable effect on the results.

\begin{figure}
    \centering
    \includegraphics[width=1\columnwidth]{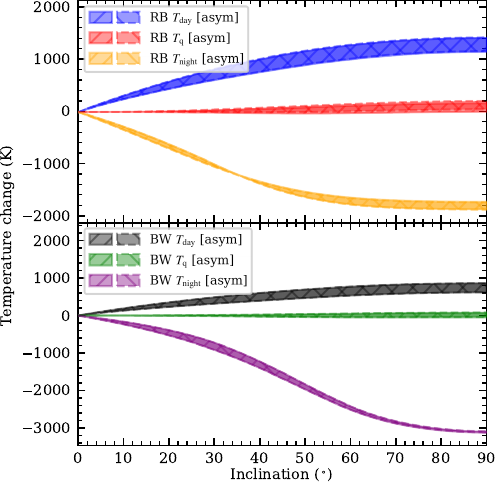}
    \caption{Simulations of redbacks (top) and black widows (bottom), with the predicted change in temperature shown as a function of inclination. The temperature change is given relative to the observed temperature at $i = 0^\circ$. Regions highlighted correspond to Roche-lobe filling factors ranging from 0.7 to 0.99. Models with asymmetric heating are indicated by dashed regions with hashing in the opposite direction, as shown in the legend -- although the difference between the symmetric and asymmetric models is subtle and the two overlap substantially.}
    \label{fig:model}
\end{figure}

The results of these simulations are presented in Figure~\ref{fig:model}. Indeed, the models clearly show that the temperature measured at quadrature remains relatively constant, regardless of inclination, while the day and night side temperatures can vary substantially. 

The greatest inclination-based temperature change at quadrature is 270\,K, for the asymmetrically heated redback at $\phi = 0.75$ (i.e. $T_\mathrm{q2}$) and a filling factor of 0.7. This can be minimized by averaging the temperatures at both points of quadrature together, giving a change of $\Delta T_\mathrm{q} = 210$\,K. This small temperature change is reduced further for systems with filling factors close to 1 -- which appears to be the case in many spiders. The primary source of the change is the phase bin over which the `observation' is performed, which biases the measurement towards phases with greater flux -- decreasing the width of the phase bin to 10 per cent of the orbit reduces the temperature change to 90\,K.

Meanwhile, both the day- and night-side temperatures depend strongly on the orbital inclination. 
The magnitude of the variation depends primarily on how irradiated the companion is -- for the strongly-irradiated redbacks, the change is between 1100--1400\,K for the day side and 1700--1900\,K for the night side, depending on the filling factor. For the black widow models, the day-side change is smaller but still important: approximately 700--800\,K, again depending on the filling factor. However, the night-side temperature changes substantially, by more than 3000\,K, regardless of the filling factor. Decreasing the phase bin width to 10 per cent of the orbit can slightly increase the change in most cases, by around 50--100\,K. For the night side of black widows, however, a narrower phase bin makes a large difference -- up to 700\,K in edge-on systems. 
As might be expected, the redback with low irradiation shows no significant changes -- its temperature is roughly constant throughout the orbit, and thus is not shown in Figure~\ref{fig:model}. 

These models clearly show that the temperature measured at quadrature $T_\mathrm{q}$ is essentially inclination-independent, to within a tolerance of less than 200\,K for a typical redback or black widow. In addition, $T_\mathrm{q}$ is less affected than $T_\mathrm{day}$ or $T_\mathrm{night}$ by the width of the phase bin used for averaging observations, particularly for nearly Roche-lobe filling systems. As such, $T_\mathrm{q}$ is primarily a function of the base temperature of the companion $T_\mathrm{base}$ and the heating luminosity $L_\mathrm{H}$, whereas $T_\mathrm{day}$ and $T_\mathrm{night}$ depend on three parameters: $T_\mathrm{base}$, $L_\mathrm{H}$, and $i$.
Therefore, in systems where irradiation is significant, independent measurements of $T_\mathrm{q}$ can provide an additional constraint on the companion to better pin down $T_\mathrm{base}$ and $L_\mathrm{H}$, both of which are crucial to accurate modelling.
Also, in systems such as J1810, where reliable night-side temperatures are challenging to obtain, measurements of $T_\mathrm{q}$ and $T_\mathrm{day}$ can provide an alternative method for constraining the orbital inclination in the absence of a well-measured $T_\mathrm{night}$.  

\subsection{Mass constraints}
\label{sec:masses}

Thanks to the availability of radio timing solutions for all three systems, the projected semi-major axis of the pulsar orbit $x_\mathrm{p}$ for each system has already been measured. This allows for the mass ratio $q = M_2 / M_\mathrm{NS} = K_1 / K_2$ to be directly determined, where $K_1 = 2 \pi c x_\mathrm{p} / P_\mathrm{b}$ is the semi-amplitude of the projected pulsar radial velocity.
Using our GTC spectroscopic results, we can now derive new constraints on the masses and orbital parameters for each of the three spiders.

\subsubsection{PSR J1048+2339}

The radial velocity fits of J1048 allow us to place strong constraints on its mass ratio. Assuming the centre-of-mass $K_2$ is well-bracketed by the observed $K_\mathrm{metals, red}$ and $K_\mathrm{metals, blue}$, then $K_2$ must be in the range $K_\mathrm{metals, red} (= 340\kms{}) < K_2 < K_\mathrm{metals, blue} (= 375\kms{})$ at a 1-$\sigma$ confidence level. Using $x_\mathrm{p} = 0.836122 \pm 0.000003$\,lt-s for the projected pulsar semi-major axis \citep{deneva16}, the 1-$\sigma$ constraints on the mass ratio become $0.194 < q < 0.214$. This refines the previous constraints of $0.209 < q < 0.250$ from \citet{zanon21} to a lower and tighter range. 
Furthermore, J1048 exhibits gamma-ray eclipses, which \citet{clark23} have leveraged to determine a minimum inclination of $i > 80.4^\circ$. Using this limit, we can directly constrain the masses of both components: the range of possible neutron star masses becomes $M_\mathrm{NS} = 1.50$--$2.04\ms{}$, and the companion mass must lie in the range $M_2 = 0.32$--$0.40\ms{}$.

\subsubsection{PSR J1810+1744}

Similarly, for J1810, the measured radial velocities allow for good constraints on the mass ratio. Using $x_\mathrm{p} = 0.095389 \pm 0.000009$\,lt-s \citep{fermi23}, and again assuming our measured $K_\mathrm{Balmer}$ and $K_\mathrm{metals}$ bracket the true $K_2$, the 1-$\sigma$ constraints become $0.027 < q < 0.033$, in agreement with those found by \citet{romani21} by optical modelling.

Unlike J1048, J1810 does not show gamma-ray eclipses. However, as gamma-rays are expected to be produced close to the neutron star, the lack of eclipses gives a maximum inclination for J1810 of $i < 84.7^\circ$ \citep{clark23}. With this, we determine a minimum neutron star mass of $1.3\ms{}$. 
We cannot place robust upper limits on the neutron star mass, as the orbital solution is not precise enough due to the lack of reliable radial velocity measurements of the night side of the companion.
Assuming a maximum neutron star mass of $2.5\ms{}$, the companion mass can be constrained to the range $M_2 = 0.04$--$0.07\ms{}$.
If instead we use $i = 66.3 \pm 0.5^\circ$, as determined by \citet{romani21}, the lower limit on the neutron star mass becomes $M_\mathrm{NS} > 1.7\ms{}$, and the companion mass is constrained to a slightly higher range: $0.05\ms{} < M_2 < 0.08\ms{}$.

\subsubsection{PSR J1908+2105}

As detailed in Section~\ref{sec:j1908}, we did not detect any significant radial velocity variations from the companion of J1908. Instead, we determined an upper limit of $K_2 < 32\kms{}$ at a confidence level of 3-$\sigma$. 
Again, J1908 already has a precisely determined radio timing solution, with $x_\mathrm{p} = 0.116895^{+0.00002}_{-0.000002}$\,lt-s \citep{deneva21}. As such, the upper limit on $K_2$ immediately provides a lower limit on the mass ratio of $q > 0.55$. This is the highest of any confirmed spider system -- only beaten by 1FGL J0523.5-2529, a candidate redback with $q = 0.61 \pm 0.06$ \citep{strader14} -- and immediately confirms J1908 to be a redback system. Even if we use the less stringent $K_2$ upper limit of $49\kms{}$ (obtained with $T_0$ left as a free parameter -- see Section~\ref{sec:vels}), the minimum mass ratio is still $0.36$ -- firmly cementing J1908 as a member of the redback population.

\begin{figure}
    \centering
    \includegraphics[width=1\columnwidth]{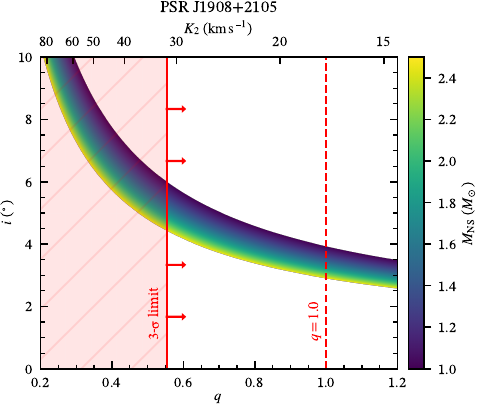}
    \caption{Mass constraints for J1908, using the radio timing of \citet{deneva21} and assuming a minimum neutron star mass of $1\ms{}$. The 3-$\sigma$ upper limit of $K_2 < 32\kms{}$, corresponding to $q > 0.55$, is indicated by the solid red line, with the parameter space it excludes highlighted to the left in red. The dashed line indicates $q = 1.0$. The coloured region indicates a range of neutron star masses from $1.0$ to $2.5\ms{}$, with lighter colours indicating higher masses.}
    \label{fig:j1908mass}
\end{figure}

Furthermore, while the neutron star mass cannot be tightly constrained until $K_2$ is measured, we can flip the problem to constrain the inclination of J1908. $M_\mathrm{NS}$ can be computed as a function of both $q$ and $i$. Then, taking the maximum neutron star mass as $2.5\ms{}$ and the \textit{minimum} neutron star mass as $1.0\ms{}$, the invalid combinations of $q$ and $i$ can be ruled out. The result is shown in Figure~\ref{fig:j1908mass} -- with this, we constrain the inclination to $i < 6.0^\circ$, making J1908 the most face-on confirmed spider observed to date \citep[see also][for a candidate transitional MSP constrained at $i=5$--$8^\circ$]{Britt17}.
This very low orbital inclination combined with its low-amplitude light curve \citep{beronya23} suggests J1908 is in fact substantially irradiated, and its weak orbital modulation is merely a product of the viewing angle. As such, we also expect the X-ray orbital modulation in J1908 to have a very small amplitude.

Such a low inclination system naturally raises questions to the nature of its radio eclipses. According to \citet{deneva21}, J1908 presents radio eclipses for approximately 40 per cent of its orbit -- typical of a redback system. The source of the radio eclipses in spider pulsars is expected to be due to a tail of ablated material swept behind the companion, which blocks the pulsars radio signal. For radio eclipses to occur via this mechanism in J1908, ablated material would have to be swept significantly out of the orbital plane; however, this would still only occult the pulsar close to its superior conjunction.

One possible explanation could be that the unusually massive companion produces a stronger wind, altering the geometry of the shock between the companion and pulsar wind. This could result in a nearly flat shock, or a shock slightly wrapped around the pulsar, which could eject and confine obscuring material high above the orbital plane.
Intrabinary shock models \citep[e.g.,][]{Wadiasingh17} could be used to constrain the shock opening angle in this special viewing geometry.

\subsection{H$\alpha$ emission variability and Ca enrichment}
\label{sec:misc}

Spider systems exhibit a variety of unique and unusual spectral features that become clear particularly when compared to template spectra of isolated main sequence stars. Some of these characteristics may be due to their dynamic environment, while others are a result of their violent evolution -- including the supernovae that produce their pulsar primaries. While high resolution, deep studies of spider companions are needed to properly evaluate and quantify these effects, the evidence can still be seen at lower resolutions.

Similarly to the observations of \citet{zanon21}, conducted exactly one month after those presented here, J1048 shows H$\alpha$ emission that peaks twice per orbit, once after superior and once after inferior conjunction (see Figure~\ref{fig:j1048ews}). However, here the emission is much stronger, reaching in excess of three times the continuum level at some points. In addition, the emission peak is broad and both red- and blue-shifted over the orbit, sometimes forming double peaks. The substantial difference in emission is somewhat surprising considering that, in both cases, simultaneous photometry shows the companion to be in an irradiated state. The H$\alpha$ emission seen here is more akin to the behaviour found by \citet{strader19} in some of their epochs which cover a couple months in 2018. We also observed the same \ion{He}{I} 5876\,\AA{} emission line seen in their spectra -- here we found it to be coincident with the peaks in H$\alpha$ emission. The common factor between all three sets of observations is the short time-scales of both line strength and width variations, which suggest the origin is related to the intrabinary shock between the companion and the pulsar, rather than an accretion disc.

The chemistry of donor stars in binary systems offers a unique view into their histories. Both accretion and ablation will strip the outer layers of the companion, modifying its observed composition. Additionally, it is possible for the supernova explosion that produces the neutron star primary to pollute the companion, increasing the abundance of $\alpha$-elements. Only two binary MSPs have been studied with sufficient detail to measure abundances: PSR J1740-5340 \citep{mucciarelli13} and PSR J1023+0038 \citep{shahbaz22}. Both show evidence of this, with increased Li and Ca abundances derived from optical spectra. 

As shown in Section~\ref{sec:j1908}, the near-infrared \ion{Ca}{II} triplet is enhanced in J1908 with respect to the template spectra, which have solar metallicities. The strength of the enhancement forced us to consider the \ion{Ca}{II} lines separately from the optical absorption lines for optimal subtraction, as all of the absorption features could not be matched simultaneously. The size of this enhancement is encapsulated in the veiling factor, $f_\mathrm{veil}$, which is $0.67\pm0.01$ for the optical lines and $1.65\pm0.03$ for the \ion{Ca}{II} triplet. This is suggestive of Ca enrichment in the companion of J1908, with the most probable source being supernova pollution. We also see a hint of enhancement in J1048, with the \ion{Na}{I} 5683+5688\,\AA{} and \ion{Ca}{I} 6169\,\AA{} lines being stronger than in the best-fitting templates. Thus, in both J1048 and J1908, we find tentative evidence of Ca and Na enrichment. However, higher resolution observations are needed to better quantify it and determine its origins.

\section{Conclusions}
\label{sec:concl}

We have obtained and thoroughly analysed GTC-OSIRIS optical spectra of the companions of three spider pulsars, including the deepest observations of PSR J1048+2339 to date, the first spectroscopic temperatures of PSR J1810+1744, and the first in-depth spectral analysis of PSR J1908+2105 throughout its orbit. The three systems we have considered cover a full range of inclinations: J1048 is edge-on ($i \gtrsim 80^\circ$), J1810 is at an intermediate inclination ($i \sim 60^\circ$), and J1908 is remarkably face-on ($i < 6^\circ$).

For all three systems, we have measured independent line-based temperatures throughout their orbits. For both redback systems, the companion effective temperature has been traced well throughout the entire orbit, showing a clear day-night contrast in J1048 (from $4690^{+51}_{-48}$\,K to $4072^{+32}_{-31}$\,K), while J1908 appears to be essentially unchanging in temperature (at $5706^{+72}_{-73}$\,K) due to its low inclination. For the black widow J1810, we have measured the day side ($T_\mathrm{day} = 7827^{+90}_{-89}$\,K) and quadrature temperatures ($T_\mathrm{q1} = 7085^{+87}_{-83}$\,K at $\phi = 0.25$, $T_\mathrm{q2} = 7113^{+138}_{-125}$\,K at $\phi = 0.75$).

We have also measured radial velocities from all three spiders, and -- using existing inclination constraints -- mass constraints have been determined. 
For J1048,  we found using different sets of temperature-sensitive absorption lines produces a change in the radial velocity curve semi-amplitude of approximately 10 per cent ($K_\mathrm{metals, red} = 344 \pm 4\kms{}$ vs. $K_\mathrm{metals, blue} = 372 \pm 3\kms{}$), similar to what was found in PSR J2215+5135 by \citet{linares18}. This allows us to greatly refine the mass constraints on this system to $0.194 < q < 0.214$,  $M_\mathrm{NS} = 1.50$--$2.04\ms{}$, and $M_2 = 0.32$--$0.40\ms{}$. 
Tentative evidence of a similar separation was also seen between Balmer and metallic lines in J1810, although its faint dark side severely limits the precision of the result ($K_\mathrm{Balmer} = 448 \pm 19\kms{}$ vs. $K_\mathrm{metals} = 491 \pm 32\kms{} $). This gives a mass ratio constraint of $0.027 < q < 0.033$, and, using $i = 66.3 \pm 0.5^\circ$ \citep{romani21}, the limits of $M_\mathrm{NS} > 1.7\ms{}$ and $0.05\ms{} < M_2 < 0.08\ms{}$ are found.

In J1908, no significant sinusoidal radial velocity curve was found -- instead, we placed an upper limit of $32\kms{}$ on the radial velocity curve semi-amplitude. As detailed in Section~\ref{sec:masses}, this allows us to constrain the mass ratio to $q > 0.55$ and the inclination to $i < 6.0^\circ$, thus revealing J1908 to be a face-on redback system. This also makes J1908 the highest mass ratio, lowest inclination spider system discovered so far.

We have found evidence of asymmetric heating, in both J1048 and J1810, in the equivalent widths of various absorption features and their dependence on orbital phase. The asymmetries found in equivalent widths follow those seen previously in photometry of the same systems \citep{yap19, zanon21, romani21}. 
Signs of enrichment were also seen in the optical spectra of the companion stars, particularly in J1908. This is interpreted as tentative evidence of supernova pollution, as previously observed in the companions of two other spider pulsars \citep{mucciarelli13, shahbaz22}. 
Finally, we explored the inclination dependence of spectroscopic temperatures using binary models, demonstrating that the companion's temperature at quadrature is essentially independent of inclination, and thus can provide additional constraints on optical light curve modelling.

It has become evident that optical photometry alone is not enough to place robust dynamical constraints on spiders. The use of focused spectroscopic studies can quantify the effects of irradiation in these systems, and leverage them to obtain robust and precise neutron star masses. Combined with optical light curve modelling, this will provide the best mass measurements of spiders, allowing us to continue to hunt down the most massive neutron stars.

\section*{Acknowledgements}

This project has received funding from the European Research Council (ERC) under the European Union's Horizon 2020 research and innovation programme (Grant agreement No. 101002352, PI: M. Linares).
Based on observations made with the Gran Telescopio Canarias (GTC), installed at the Spanish Observatorio del Roque de los Muchachos of the Instituto de Astrof\'isica de Canarias, on the island of La Palma.
The Starlink software \citet{starlink} is currently supported by the East Asian Observatory.
JC acknowledges support by the Spanish Ministry of Science via the Plan de Generaci\'on de Conocimiento under grant PID2022-143331NB-I00.
The authors thank the late T. Marsh for the use of the \textsc{pamela} and \textsc{molly} packages, E. Parent for discussions on the radio eclipses in J1908, and K. Conroy and D. Jones for discussion and advice on simulating observations in PHOEBE.

\section*{Data Availability}

All data used in this work is publicly accessible through the GTC Public Archive (\href{https://gtc.sdc.cab.inta-csic.es/gtc/}{gtc.sdc.cab.inta-csic.es/gtc/}). The three datasets can be found under the GTC Program IDs GTC106-19B, GTC66-15A, and GTC77-20A for observations of PSR J1048+2339, PSR J1810+1744, and PSR J1908+2105, respectively.



\bibliographystyle{mnras}
\bibliography{gtc-pulsars.bib}

\begin{thebibliography}{}
\makeatletter
\relax
\def\mn@urlcharsother{\let\do\@makeother \do\$\do\&\do\#\do\^\do\_\do\%\do\~}
\def\mn@doi{\begingroup\mn@urlcharsother \@ifnextchar [ {\mn@doi@} {\mn@doi@[]}}
\def\mn@doi@[#1]#2{\def\@tempa{#1}\ifx\@tempa\@empty \href {http://dx.doi.org/#2} {doi:#2}\else \href {http://dx.doi.org/#2} {#1}\fi \endgroup}
\def\mn@eprint#1#2{\mn@eprint@#1:#2::\@nil}
\def\mn@eprint@arXiv#1{\href {http://arxiv.org/abs/#1} {{\tt arXiv:#1}}}
\def\mn@eprint@dblp#1{\href {http://dblp.uni-trier.de/rec/bibtex/#1.xml} {dblp:#1}}
\def\mn@eprint@#1:#2:#3:#4\@nil{\def\@tempa {#1}\def\@tempb {#2}\def\@tempc {#3}\ifx \@tempc \@empty \let \@tempc \@tempb \let \@tempb \@tempa \fi \ifx \@tempb \@empty \def\@tempb {arXiv}\fi \@ifundefined {mn@eprint@\@tempb}{\@tempb:\@tempc}{\expandafter \expandafter \csname mn@eprint@\@tempb\endcsname \expandafter{\@tempc}}}

\bibitem[\protect\citeauthoryear{{Allard}, {Homeier}  \& {Freytag}}{{Allard} et~al.}{2011}]{allard11}
{Allard} F.,  {Homeier} D.,   {Freytag} B.,  2011, in {Johns-Krull} C.,  {Browning} M.~K.,   {West} A.~A.,  eds,  Astronomical Society of the Pacific Conference Series Vol. 448, 16th Cambridge Workshop on Cool Stars, Stellar Systems, and the Sun. p.~91 (\mn@eprint {arXiv} {1011.5405}), \mn@doi{10.48550/arXiv.1011.5405}

\bibitem[\protect\citeauthoryear{{Bagnulo}, {Jehin}, {Ledoux}, {Cabanac}, {Melo}, {Gilmozzi}  \& {ESO Paranal Science Operations Team}}{{Bagnulo} et~al.}{2003}]{uves}
{Bagnulo} S.,  {Jehin} E.,  {Ledoux} C.,  {Cabanac} R.,  {Melo} C.,  {Gilmozzi} R.,   {ESO Paranal Science Operations Team} 2003, The Messenger, \href {https://ui.adsabs.harvard.edu/abs/2003Msngr.114...10B} {114, 10}

\bibitem[\protect\citeauthoryear{{Beronya}, {Kirichenko}, {Zharikov}, {Karpova}, {Zyuzin}  \& {Shibanov}}{{Beronya} et~al.}{2023}]{beronya23}
{Beronya} D.~M.,  {Kirichenko} A.,  {Zharikov} S.~V.,  {Karpova} A.~V.,  {Zyuzin} D.,   {Shibanov} Y.~A.,  2023, \mn@doi [Zhurnal Teknicheskoy Fiziki] {http://dx.doi.org/10.61011/JTF.2023.12.56826.f236-23}, 92, 1803

\bibitem[\protect\citeauthoryear{{Breton} et~al.,}{{Breton} et~al.}{2013}]{breton13}
{Breton} R.~P.,  et~al., 2013, \mn@doi [\apj] {10.1088/0004-637X/769/2/108}, \href {https://ui.adsabs.harvard.edu/abs/2013ApJ...769..108B} {769, 108}

\bibitem[\protect\citeauthoryear{{Britt}, {Strader}, {Chomiuk}, {Tremou}, {Peacock}, {Halpern}  \& {Salinas}}{{Britt} et~al.}{2017}]{Britt17}
{Britt} C.~T.,  {Strader} J.,  {Chomiuk} L.,  {Tremou} E.,  {Peacock} M.,  {Halpern} J.,   {Salinas} R.,  2017, \mn@doi [\apj] {10.3847/1538-4357/aa8e41}, \href {https://ui.adsabs.harvard.edu/abs/2017ApJ...849...21B} {849, 21}

\bibitem[\protect\citeauthoryear{{Cho}, {Halpern}  \& {Bogdanov}}{{Cho} et~al.}{2018}]{cho18}
{Cho} P.~B.,  {Halpern} J.~P.,   {Bogdanov} S.,  2018, \mn@doi [\apj] {10.3847/1538-4357/aade92}, \href {https://ui.adsabs.harvard.edu/abs/2018ApJ...866...71C} {866, 71}

\bibitem[\protect\citeauthoryear{{Clark} et~al.,}{{Clark} et~al.}{2023}]{clark23}
{Clark} C.~J.,  et~al., 2023, \mn@doi [Nature Astronomy] {10.1038/s41550-022-01874-x}, \href {https://ui.adsabs.harvard.edu/abs/2023NatAs...7..451C} {7, 451}

\bibitem[\protect\citeauthoryear{{Conroy} et~al.,}{{Conroy} et~al.}{2020}]{phoebe2.3}
{Conroy} K.~E.,  et~al., 2020, \mn@doi [\apjs] {10.3847/1538-4365/abb4e2}, \href {https://ui.adsabs.harvard.edu/abs/2020ApJS..250...34C} {250, 34}

\bibitem[\protect\citeauthoryear{{Cretignier}, {Francfort}, {Dumusque}, {Allart}  \& {Pepe}}{{Cretignier} et~al.}{2020}]{rassine}
{Cretignier} M.,  {Francfort} J.,  {Dumusque} X.,  {Allart} R.,   {Pepe} F.,  2020, \mn@doi [\aap] {10.1051/0004-6361/202037722}, \href {https://ui.adsabs.harvard.edu/abs/2020A&A...640A..42C} {640, A42}

\bibitem[\protect\citeauthoryear{{Cromartie} et~al.,}{{Cromartie} et~al.}{2016}]{cromartie16}
{Cromartie} H.~T.,  et~al., 2016, \mn@doi [\apj] {10.3847/0004-637X/819/1/34}, \href {https://ui.adsabs.harvard.edu/abs/2016ApJ...819...34C} {819, 34}

\bibitem[\protect\citeauthoryear{{Currie}, {Berry}, {Jenness}, {Gibb}, {Bell}  \& {Draper}}{{Currie} et~al.}{2014}]{starlink}
{Currie} M.~J.,  {Berry} D.~S.,  {Jenness} T.,  {Gibb} A.~G.,  {Bell} G.~S.,   {Draper} P.~W.,  2014, in {Manset} N.,  {Forshay} P.,  eds,  Astronomical Society of the Pacific Conference Series Vol. 485, Astronomical Data Analysis Software and Systems XXIII. p.~391

\bibitem[\protect\citeauthoryear{{Deneva} et~al.,}{{Deneva} et~al.}{2016}]{deneva16}
{Deneva} J.~S.,  et~al., 2016, \mn@doi [\apj] {10.3847/0004-637X/823/2/105}, \href {https://ui.adsabs.harvard.edu/abs/2016ApJ...823..105D} {823, 105}

\bibitem[\protect\citeauthoryear{{Deneva} et~al.,}{{Deneva} et~al.}{2021}]{deneva21}
{Deneva} J.~S.,  et~al., 2021, \mn@doi [\apj] {10.3847/1538-4357/abd7a1}, \href {https://ui.adsabs.harvard.edu/abs/2021ApJ...909....6D} {909, 6}

\bibitem[\protect\citeauthoryear{{Dodge} et~al.,}{{Dodge} et~al.}{2024}]{dodge24}
{Dodge} O.~G.,  et~al., 2024, \mn@doi [\mnras] {10.1093/mnras/stae211}, \href {https://ui.adsabs.harvard.edu/abs/2024MNRAS.528.4337D} {528, 4337}

\bibitem[\protect\citeauthoryear{{Hessels} et~al.,}{{Hessels} et~al.}{2011}]{hessels11}
{Hessels} J.~W.~T.,  et~al., 2011, in {Burgay} M.,  {D'Amico} N.,  {Esposito} P.,  {Pellizzoni} A.,   {Possenti} A.,  eds,  American Institute of Physics Conference Series Vol. 1357, Radio Pulsars: An Astrophysical Key to Unlock the Secrets of the Universe. pp 40--43 (\mn@eprint {arXiv} {1101.1742}), \mn@doi{10.1063/1.3615072}

\bibitem[\protect\citeauthoryear{{Hinkle}, {Wallace}, {Valenti}  \& {Harmer}}{{Hinkle} et~al.}{2000}]{hinkle00}
{Hinkle} K.,  {Wallace} L.,  {Valenti} J.,   {Harmer} D.,  2000, {Visible and Near Infrared Atlas of the Arcturus Spectrum 3727-9300 A}

\bibitem[\protect\citeauthoryear{{Husser}, {Wende-von Berg}, {Dreizler}, {Homeier}, {Reiners}, {Barman}  \& {Hauschildt}}{{Husser} et~al.}{2013}]{phoenix}
{Husser} T.~O.,  {Wende-von Berg} S.,  {Dreizler} S.,  {Homeier} D.,  {Reiners} A.,  {Barman} T.,   {Hauschildt} P.~H.,  2013, \mn@doi [\aap] {10.1051/0004-6361/201219058}, \href {https://ui.adsabs.harvard.edu/abs/2013A&A...553A...6H} {553, A6}

\bibitem[\protect\citeauthoryear{{Kandel} \& {Romani}}{{Kandel} \& {Romani}}{2023}]{kandel23}
{Kandel} D.,  {Romani} R.~W.,  2023, \mn@doi [\apj] {10.3847/1538-4357/aca524}, \href {https://ui.adsabs.harvard.edu/abs/2023ApJ...942....6K} {942, 6}

\bibitem[\protect\citeauthoryear{{Linares}}{{Linares}}{2020}]{linares20}
{Linares} M.,  2020, in Multifrequency Behaviour of High Energy Cosmic Sources - XIII. 3-8 June 2019. Palermo. p.~23 (\mn@eprint {arXiv} {1910.09572}), \mn@doi{10.22323/1.362.0023}

\bibitem[\protect\citeauthoryear{{Linares}, {Shahbaz}  \& {Casares}}{{Linares} et~al.}{2018}]{linares18}
{Linares} M.,  {Shahbaz} T.,   {Casares} J.,  2018, \mn@doi [\apj] {10.3847/1538-4357/aabde6}, \href {https://ui.adsabs.harvard.edu/abs/2018ApJ...859...54L} {859, 54}

\bibitem[\protect\citeauthoryear{{Marsh}}{{Marsh}}{1989}]{marsh89}
{Marsh} T.~R.,  1989, \mn@doi [\pasp] {10.1086/132570}, \href {https://ui.adsabs.harvard.edu/abs/1989PASP..101.1032M} {101, 1032}

\bibitem[\protect\citeauthoryear{{Marsh}, {Robinson}  \& {Wood}}{{Marsh} et~al.}{1994}]{marsh94}
{Marsh} T.~R.,  {Robinson} E.~L.,   {Wood} J.~H.,  1994, \mn@doi [\mnras] {10.1093/mnras/266.1.137}, \href {https://ui.adsabs.harvard.edu/abs/1994MNRAS.266..137M} {266, 137}

\bibitem[\protect\citeauthoryear{{Miraval Zanon} et~al.,}{{Miraval Zanon} et~al.}{2021}]{zanon21}
{Miraval Zanon} A.,  et~al., 2021, \mn@doi [\aap] {10.1051/0004-6361/202040071}, \href {https://ui.adsabs.harvard.edu/abs/2021A&A...649A.120M} {649, A120}

\bibitem[\protect\citeauthoryear{{Mucciarelli}, {Salaris}, {Lanzoni}, {Pallanca}, {Dalessandro}  \& {Ferraro}}{{Mucciarelli} et~al.}{2013}]{mucciarelli13}
{Mucciarelli} A.,  {Salaris} M.,  {Lanzoni} B.,  {Pallanca} C.,  {Dalessandro} E.,   {Ferraro} F.~R.,  2013, \mn@doi [\apjl] {10.1088/2041-8205/772/2/L27}, \href {https://ui.adsabs.harvard.edu/abs/2013ApJ...772L..27M} {772, L27}

\bibitem[\protect\citeauthoryear{{Polzin} et~al.,}{{Polzin} et~al.}{2018}]{polzin18}
{Polzin} E.~J.,  et~al., 2018, \mn@doi [\mnras] {10.1093/mnras/sty349}, \href {https://ui.adsabs.harvard.edu/abs/2018MNRAS.476.1968P} {476, 1968}

\bibitem[\protect\citeauthoryear{{Pr{\v{s}}a} et~al.,}{{Pr{\v{s}}a} et~al.}{2016}]{phoebe2}
{Pr{\v{s}}a} A.,  et~al., 2016, \mn@doi [\apjs] {10.3847/1538-4365/227/2/29}, \href {https://ui.adsabs.harvard.edu/abs/2016ApJS..227...29P} {227, 29}

\bibitem[\protect\citeauthoryear{{Romani} \& {Sanchez}}{{Romani} \& {Sanchez}}{2016}]{romani16}
{Romani} R.~W.,  {Sanchez} N.,  2016, \mn@doi [\apj] {10.3847/0004-637X/828/1/7}, \href {https://ui.adsabs.harvard.edu/abs/2016ApJ...828....7R} {828, 7}

\bibitem[\protect\citeauthoryear{{Romani}, {Kandel}, {Filippenko}, {Brink}  \& {Zheng}}{{Romani} et~al.}{2021}]{romani21}
{Romani} R.~W.,  {Kandel} D.,  {Filippenko} A.~V.,  {Brink} T.~G.,   {Zheng} W.,  2021, \mn@doi [\apjl] {10.3847/2041-8213/abe2b4}, \href {https://ui.adsabs.harvard.edu/abs/2021ApJ...908L..46R} {908, L46}

\bibitem[\protect\citeauthoryear{{Romani}, {Kandel}, {Filippenko}, {Brink}  \& {Zheng}}{{Romani} et~al.}{2022}]{romani22}
{Romani} R.~W.,  {Kandel} D.,  {Filippenko} A.~V.,  {Brink} T.~G.,   {Zheng} W.,  2022, \mn@doi [\apjl] {10.3847/2041-8213/ac8007}, \href {https://ui.adsabs.harvard.edu/abs/2022ApJ...934L..17R} {934, L17}

\bibitem[\protect\citeauthoryear{{Schroeder} \& {Halpern}}{{Schroeder} \& {Halpern}}{2014}]{schroeder14}
{Schroeder} J.,  {Halpern} J.,  2014, \mn@doi [\apj] {10.1088/0004-637X/793/2/78}, \href {https://ui.adsabs.harvard.edu/abs/2014ApJ...793...78S} {793, 78}

\bibitem[\protect\citeauthoryear{{Shahbaz}, {Gonz{\'a}lez-Hern{\'a}ndez}, {Breton}, {Kennedy}, {Mata S{\'a}nchez}  \& {Linares}}{{Shahbaz} et~al.}{2022}]{shahbaz22}
{Shahbaz} T.,  {Gonz{\'a}lez-Hern{\'a}ndez} J.~I.,  {Breton} R.~P.,  {Kennedy} M.~R.,  {Mata S{\'a}nchez} D.,   {Linares} M.,  2022, \mn@doi [\mnras] {10.1093/mnras/stac492}, \href {https://ui.adsabs.harvard.edu/abs/2022MNRAS.513...71S} {513, 71}

\bibitem[\protect\citeauthoryear{{Smith} et~al.,}{{Smith} et~al.}{2023a}]{fermi23}
{Smith} D.~A.,  et~al., 2023a, \mn@doi [\apj] {10.3847/1538-4357/acee67}, \href {https://ui.adsabs.harvard.edu/abs/2023ApJ...958..191S} {958, 191}

\bibitem[\protect\citeauthoryear{{Smith} et~al.,}{{Smith} et~al.}{2023b}]{smith23}
{Smith} D.~A.,  et~al., 2023b, \mn@doi [\apj] {10.3847/1538-4357/acee67}, \href {https://ui.adsabs.harvard.edu/abs/2023ApJ...958..191S} {958, 191}

\bibitem[\protect\citeauthoryear{{Strader}, {Chomiuk}, {Sonbas}, {Sokolovsky}, {Sand}, {Moskvitin}  \& {Cheung}}{{Strader} et~al.}{2014}]{strader14}
{Strader} J.,  {Chomiuk} L.,  {Sonbas} E.,  {Sokolovsky} K.,  {Sand} D.~J.,  {Moskvitin} A.~S.,   {Cheung} C.~C.,  2014, \mn@doi [\apjl] {10.1088/2041-8205/788/2/L27}, \href {https://ui.adsabs.harvard.edu/abs/2014ApJ...788L..27S} {788, L27}

\bibitem[\protect\citeauthoryear{{Strader} et~al.,}{{Strader} et~al.}{2019}]{strader19}
{Strader} J.,  et~al., 2019, \mn@doi [\apj] {10.3847/1538-4357/aafbaa}, \href {https://ui.adsabs.harvard.edu/abs/2019ApJ...872...42S} {872, 42}

\bibitem[\protect\citeauthoryear{{Tidemann}}{{Tidemann}}{2023}]{tidemann23}
{Tidemann} A.,  2023, Master's thesis, NTNU, Trondheim, NO

\bibitem[\protect\citeauthoryear{{Turchetta}, {Linares}, {Koljonen}  \& {Sen}}{{Turchetta} et~al.}{2023}]{turchetta23}
{Turchetta} M.,  {Linares} M.,  {Koljonen} K.,   {Sen} B.,  2023, \mn@doi [\mnras] {10.1093/mnras/stad2435}, \href {https://ui.adsabs.harvard.edu/abs/2023MNRAS.525.2565T} {525, 2565}

\bibitem[\protect\citeauthoryear{Virtanen et~al.,}{Virtanen et~al.}{2020}]{scipy}
Virtanen P.,  et~al., 2020, \mn@doi [Nature Methods] {10.1038/s41592-019-0686-2}, \href {https://rdcu.be/b08Wh} {17, 261}

\bibitem[\protect\citeauthoryear{{Voisin}, {Kennedy}, {Breton}, {Clark}  \& {Mata-S{\'a}nchez}}{{Voisin} et~al.}{2020}]{voisin20}
{Voisin} G.,  {Kennedy} M.~R.,  {Breton} R.~P.,  {Clark} C.~J.,   {Mata-S{\'a}nchez} D.,  2020, \mn@doi [\mnras] {10.1093/mnras/staa2876}, \href {https://ui.adsabs.harvard.edu/abs/2020MNRAS.499.1758V} {499, 1758}

\bibitem[\protect\citeauthoryear{{Wade} \& {Horne}}{{Wade} \& {Horne}}{1988}]{wade88}
{Wade} R.~A.,  {Horne} K.,  1988, \mn@doi [\apj] {10.1086/165905}, \href {https://ui.adsabs.harvard.edu/abs/1988ApJ...324..411W} {324, 411}

\bibitem[\protect\citeauthoryear{{Wadiasingh}, {Harding}, {Venter}, {B{\"o}ttcher}  \& {Baring}}{{Wadiasingh} et~al.}{2017}]{Wadiasingh17}
{Wadiasingh} Z.,  {Harding} A.~K.,  {Venter} C.,  {B{\"o}ttcher} M.,   {Baring} M.~G.,  2017, \mn@doi [\apj] {10.3847/1538-4357/aa69bf}, \href {https://ui.adsabs.harvard.edu/abs/2017ApJ...839...80W} {839, 80}

\bibitem[\protect\citeauthoryear{{Yao}, {Manchester}  \& {Wang}}{{Yao} et~al.}{2017}]{yao17}
{Yao} J.~M.,  {Manchester} R.~N.,   {Wang} N.,  2017, \mn@doi [\apj] {10.3847/1538-4357/835/1/29}, \href {https://ui.adsabs.harvard.edu/abs/2017ApJ...835...29Y} {835, 29}

\bibitem[\protect\citeauthoryear{{Yap}, {Li}, {Kong}, {Takata}, {Lee}  \& {Hui}}{{Yap} et~al.}{2019}]{yap19}
{Yap} Y.~X.,  {Li} K.~L.,  {Kong} A.~K.~H.,  {Takata} J.,  {Lee} J.,   {Hui} C.~Y.,  2019, \mn@doi [\aap] {10.1051/0004-6361/201834545}, \href {https://ui.adsabs.harvard.edu/abs/2019A&A...621L...9Y} {621, L9}

\makeatother
\end{thebibliography}



\appendix

\section{BT-Settl template spectra}
\label{sec:templates}

As part of this analysis, 60 BT-Settl AGS2009 synthetic spectra \citep{allard11} were rebinned, cropped, calibrated and normalised. This library of synthetic spectra cover the wavelength range 2500--20000\,\AA{} at a resolution of $\sim$0.05\,\AA{} px$^{-1}$ in the optical wavelengths and $\sim$0.15\,\AA{} px$^{-1}$ in the near-infrared. The spectra span a model parameter space of $T_\mathrm{eff}$ = 2600--10000\,K, with $\log(g)=4.5$ and solar metallicity.

To produce the template spectra used in the analysis detailed in Sections~\ref{sec:rvs} and \ref{sec:optsub}, these full-resolution synthetic spectra were artificially broadened to match the instrumental resolution of each set of observations, and then rebinned to a uniform velocity scale matching that of the corresponding set of observations. The \texttt{norma} package\footnote{\url{http://github.com/jsimpson-astro/norma}}, based partially on the algorithm from RASSINE \citep{rassine}, was used to normalise each set of templates, both before and after broadening.

\begin{figure*}
    \centering
    \includegraphics[width=0.33\textwidth]{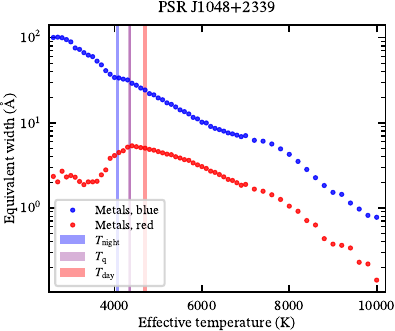}
    \includegraphics[width=0.33\textwidth]{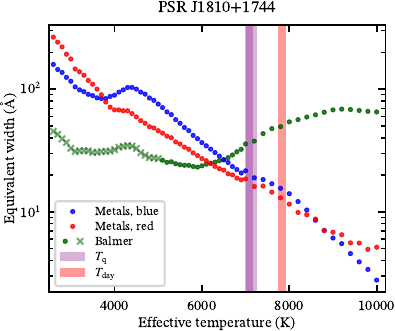}
    \includegraphics[width=0.33\textwidth]{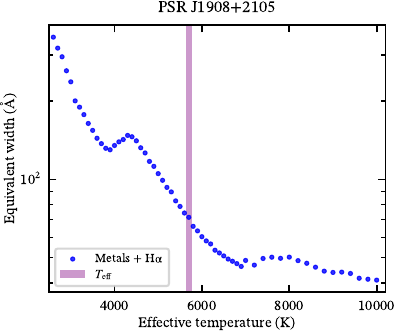}
    \caption{Equivalent widths of BT-Settl templates with the masks used for each system. Measured temperatures from optimal subtraction are indicated by vertical bars. For the equivalent widths measured from the Balmer series in J1810, crosses indicate the point where Balmer absorption lines blend with metallic lines (particularly H$\gamma$ and H$\beta$), and thus the measured equivalent widths are no longer representative of only Balmer absorption.}
    \label{fig:temp_ews}
\end{figure*}

The equivalent widths measured from these template spectra, using the masks shown in Figure~\ref{fig:j1048}, ~\ref{fig:j1810}, and~\ref{fig:j1908}, are shown in Figure~\ref{fig:temp_ews}. The temperatures determined from optimal subtraction are also highlighted, to show the sensitivity of these masks over the range of temperatures observed from each companion.

\begin{figure*}
    \centering
    \includegraphics[height=0.96\textheight]{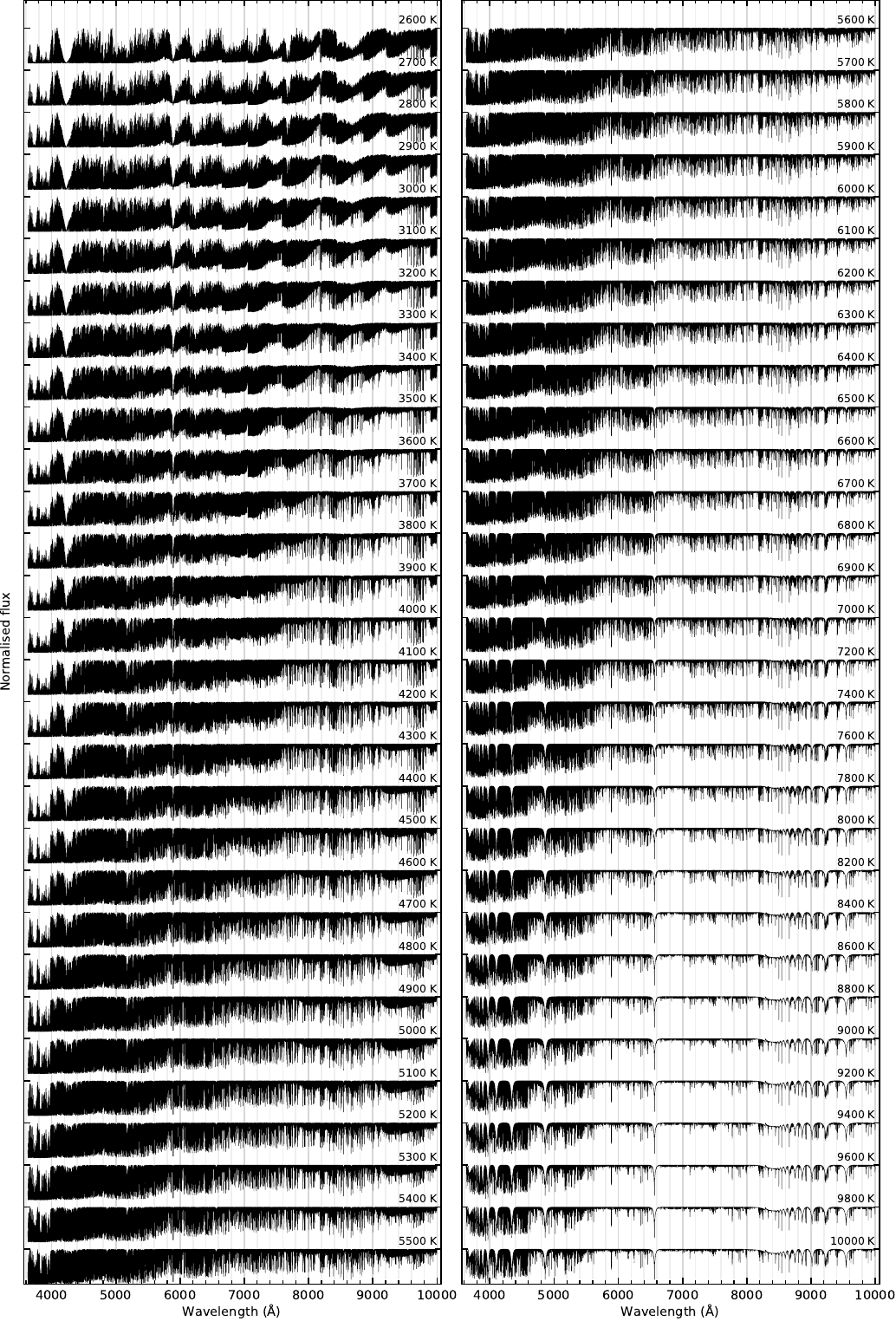}
    \caption{BT-Settl normalised spectra at full resolution, cropped to cover the wavelength range 3600--10000\,\AA{}. The effective temperature of each spectrum is labelled above it.}
    \label{fig:bt}
\end{figure*}

The full-resolution normalised synthetic spectra are shown in Figure~\ref{fig:bt}, with the corresponding effective temperature indicated. The spectra have been cropped to the wavelength range 3600--10000\,\AA{} for display purposes.

\bsp	
\label{lastpage}
\end{document}